# To tune or not to tune, a case study of ridge logistic regression in small or sparse datasets

Hana Šinkovec[1], Georg Heinze[1], Rok Blagus[2], Angelika Geroldinger[1]

[1] Section for Clinical Biometrics; Center for Medical Statistics, Informatics and Intelligent Systems; Medical University of Vienna, Austria

[2] Institute for Biostatistics and Medical Informatics, University of Ljubljana, Slovenia

**Abstract**

For finite samples with binary outcomes penalized logistic regression such as ridge logistic regression (RR) has the potential of achieving smaller mean squared errors (MSE) of coefficients and predictions than maximum likelihood estimation. There is evidence, however, that RR is sensitive to small or sparse data situations, yielding poor performance in individual datasets. In this paper, we elaborate this issue further by performing a comprehensive simulation study, investigating the performance of RR in comparison to Firth's correction that has been shown to perform well in low-dimensional settings. Performance of RR strongly depends on the choice of complexity parameter that is usually tuned by minimizing some measure of the out-of-sample prediction error or information criterion. Alternatively, it may be determined according to prior assumptions about true effects. As shown in our simulation and illustrated by a data example, values optimized in small or sparse datasets are negatively correlated with optimal values and suffer from substantial variability which translates into large MSE of coefficients and large variability of calibration slopes. In contrast, if the degree of shrinkage is pre-specified, accurate coefficients and predictions can be obtained even in non-ideal settings such as encountered in the context of rare outcomes or sparse predictors.

# 1  Background

In medical research, logistic regression is commonly used to study the relationship between a binary outcome and a set of covariates. For a dataset with a balanced outcome and sufficient sample size, the maximum likelihood (ML) estimation of the regression coefficients facilitates inference, i.e. interpretability of effect estimates, as well as accuracy of predictions given the covariates. Thus, ML logistic regression may be used for explanation or prediction, depending on context. These attractive properties of the ML logistic regression, however, vanish when the sample size is small or the prevalence of one of the two outcome levels (for some combination of exposure) is low, yielding coefficient estimates biased away from zero and very unstable predictions. In such small or sparse data settings a large number of covariates that could potentially both reduce bias and improve the accuracy of predictions even amplifies these effects instead, producing inaccurate coefficient estimates and predictions that generalize poorly on a new dataset from the same population [1, 2].

A straightforward approach to deal with the problem is to apply penalized ML logistic regression: a penalty term that is added to the log likelihood function provides shrinkage of the coefficients towards zero, hereby decreasing the variance of the ML estimates and stabilizing the predictions by pulling them towards the observed event rate [3]. A common way of shrinkage is by ridge logistic regression (RR) where the penalty is defined as minus the square of the Euclidean norm of the coefficients multiplied by a non-negative complexity parameter $\lambda$. The multiplier $\lambda$ controls the strength of the penalty, i.e. amount of shrinkage towards zero. Generally, the goal based on the idea of the bias-variance trade-off, according to which the expected prediction error can be decomposed into the three components bias, variance and irreducible error, is to find the value of $\lambda$ that balances the model between underfitting and overfitting, thus producing results that are generalizable to the underlying population [4, 5]. As compared to ML the resulting coefficients may achieve lower mean squared errors (MSE) but are usually biased towards zero, therefore conventional inference by hypothesis tests and confidence intervals based on standard errors is difficult [6]. A further complication for inference arises from the estimation of $\lambda$, which is often performed on the same data set by cross-validation (CV), as its sampling variability contributes to the uncertainty in the regression coefficients.

Tuned RR has been extensively investigated in simulation studies and was commonly found to perform well for low dimensional settings in terms of small MSE of coefficients and predictions [2, 7, 8]. However, there has been evidence that RR is sensitive to small or sparse data situations, yielding poor performance in individual datasets [9-11]. Recent recommendations, therefore, advise caution when using RR for developing prediction models in case of low sample size or low events per variable ratio and call for more research investigating the impact of specific combinations of shrinkage and tuning methods [10]. While in theory there always exists some value of $\lambda$ for which RR outperforms ML in terms of the MSE of predictions [12], choosing $\lambda$ adequately in datasets that suffer from large random sampling variation is difficult as tuning procedures based on out-of-sample prediction performance might fail to approximate the U-shaped curve arising from the bias-variance trade-off and result in an arbitrary choice of $\lambda$ that either equals the smallest or the largest value of the pre-specified range of values, yielding large variability of tuned solutions and consequently, very unstable estimates.

In the present paper we investigate the performance of different commonly used approaches to tune RR in a low-dimensional sparse data setting by means of a simulation study. We also include RR with pre-specified $\lambda$, which is interpretable as Bayesian analysis with a normal prior centered at zero [1, 13], and Firth's correction [14] (FC) in our comparison, as these approaches were proposed for similar settings [7, 10, 11, 15] and do not suffer from the convergence issues that may occur in ML estimation [16]. We structured the paper accordingly: in the following section we introduce FC and RR and describe different ways to choose the complexity parameter $\lambda$ in RR. We then demonstrate the problem of tuning in sparse data situations by an illustrative example. Subsequently, we present the setup and report the results from our simulation study with respect to the accuracy of coefficients and predictions. Finally, we summarize our main findings.

## 2 Methods

Let $y_i \in \{0,1\}, i = 1, \ldots N$, be a realization of a binary outcome variable $Y$, where $y_i = 1$ denotes an event occurring in the $i$-th observation with probability $\pi_i$. The logistic regression model associates $y_i$ to a set of corresponding covariate values $x_i = (1, x_{i1}, \ldots, x_{iK}), K < N$, by assuming

$$\pi_i = P(Y = 1|x_i) = \frac{1}{1 + \exp(-\beta_0 - \beta_1 x_{i1} - \cdots - \beta_K x_{iK})},$$

where $\beta_0$ is an intercept and $\beta_k, k = 1, \ldots, K$, are regression coefficients. The parameters $\boldsymbol{\beta} = (\beta_0, \beta_1, \ldots \beta_K)$ of the model can be obtained by the ML method, maximizing the log-likelihood function

$$\ell(\boldsymbol{\beta}) = \sum_{i=1}^{N} (y_i \log \pi_i + (1 - y_i) \log(1 - \pi_i)),$$

using an iterative algorithm [17].

### 2.1 Firth's correction

ML is asymptotically unbiased, however, in situations when data are small or sparse coefficient estimates become biased away from zero and very unstable or may even not exist [16]. To reduce the bias of ML estimates, Firth [14] proposed to penalize the likelihood function by Jeffreys' invariant prior so that the penalized log-likelihood becomes

$$\ell^*(\boldsymbol{\beta}) = \ell(\boldsymbol{\beta}) + \frac{1}{2} \log|I(\boldsymbol{\beta})|,$$

where $I(\boldsymbol{\beta})$ is the Fisher information matrix evaluated at $\boldsymbol{\beta}$. Since the intercept is included in the penalty term, the average predicted probability may not equal the observed event rate but is instead biased towards one-half. To correct for this bias that may become especially apparent in situations with unbalanced outcome, Puhr et al. [7] proposed a simple modification, FC with intercept-correction (FLIC) that alters the intercept such that average predicted probabilities become equal to the observed event rate.

## 2.2 Ridge regression

In RR coefficients are constrained by the square of the Euclidean norm of the coefficients, i.e. the penalized log-likelihood reads

$$\ell^*(\boldsymbol{\beta}) = \ell(\boldsymbol{\beta}) - \frac{\lambda}{2}\sum_{k=1}^{K}\beta_k^2,$$

where complexity parameter $\lambda, \lambda > 0$, controls the amount of shrinkage towards zero. The intercept $\beta_0$ is excluded from the penalty term, yielding an average predicted probability equal to the observed event rate. Unlike FC, RR is not invariant to linear transformation of the design matrix. Therefore, to facilitate interpretation and ensure that coefficients are represented on the same scale, suitable standardization of covariates is required, usually to zero mean and unit variance.

### 2.2.1 Tuning procedures

To select the complexity parameter $\lambda$, generally, a sequence of $\lambda$ values is pre-specified and the corresponding set of models is evaluated. The optimized $\lambda^*$ is the one that produces the model minimizing the expected out-of-sample prediction error, often estimated by CV. The out-of-sample prediction error may be defined in different ways [3, 18-20], e.g. as

- minus twice log-likelihood error, i.e. deviance ($D$) [3]

$$D = -2\sum_{i=1}^{N}(y_i \log \hat{\pi}_{(-i)} + (1-y_i)\log(1-\hat{\pi}_{(-i)})),$$

- generalized cross-validation ($GCV$) [18, 20]

$$GCV = \frac{N \cdot D}{(N - df_e)^2},$$

where $df_e$ are the effective degrees of freedom, $df_e = \text{trace}(\frac{\partial^2 \ell}{\partial^2 \boldsymbol{\beta}}(\hat{\boldsymbol{\beta}})\left(\frac{\partial^2 \ell^*}{\partial^2 \boldsymbol{\beta}}(\hat{\boldsymbol{\beta}})\right)^{-1})$,

- classification error ($CE$) [3]

$$CE = \frac{1}{N}\sum_{i=1}^{N}(y_i I(\hat{\pi}_{(-i)} < c) + (1-y_i)I(\hat{\pi}_{(-i)} > c) + \frac{1}{2}I(\hat{\pi}_{(-i)} = c)),$$

with $I$ denoting an indicator function and $c$ some cut-off, usually set to $1/2$. Since in datasets with unbalanced outcomes $c = 1/2$ would assign most of observations to the more frequent outcome level Blagus and Lusa [9] advised to set $c$ equal to the marginal event rate instead.

In the definitions above $\hat{\pi}_{(-i)}$ is the event probability estimate for the $i$-th observation computed from the model where that observation has been left out from estimation of the model parameters. Alternatively, 10-fold CV may be used to speed-up computations, however, this produces different optimized $\lambda^*$ values for different combinations of fold assignments to observations. To stabilize the selection of $\lambda^*$, 10-fold CV may be repeated several times, and a particular qunatile $\theta$ of the values obtained may be used [2, 21].

Alternatively, to avoid resampling, $\lambda$ may be tuned by using the Akaike's information criterion (AIC) [6, 22], where

$$\text{AIC} = -2\ell(\widehat{\boldsymbol{\beta}}) + 2df_e.$$

### 2.2.2 Pre-specifying the degree of shrinkage

Mathematically, RR is identical to Bayesian analysis with zero-centered univariate normal priors imposed on the coefficients [1]. The variance $v_{prior}$ of these priors is inversely proportional to $\lambda$. If the priors for the coefficients are assumed to have different variances, this translates into a penalty equal to the weighted sum of squared coefficients with different weights for each coefficient. In the Bayesian analysis approach suggested by Sullivan and Greenland the degree of shrinkage is not determined by tuning but is instead based on some prior assumptions about covariates' odds ratios that can be easily converted into $v_{prior}$ [13]. The prior variance $v_{prior}$ can be obtained from a plausible (usually 95%) prior interval for a covariate's odds ratio that has to be specified according to some background assumptions. In a particular setting Sullivan and Greenland [13] considered as plausible the 95% odds ratio interval ranging from 0.25 to 4 which translates to $v_{prior} = 1/2$. However, if one wishes to avoid the effort of specifying prior distributions applying weakly informative priors, e.g. assuming the 95% probability that the odds ratio falls between 1/16 to 16, would still be beneficial to stabilize estimates.

## 3 Illustrative example

Consider the two datasets described below, each of sample size 100 with one binary predictor $X$.

Dataset 1

|       |   | y  |   |
|-------|---|----|---|
|       |   | 0  | 1 |
| $x_1$ | 0 | 20 | 0 |
|       | 1 | 71 | 9 |

Dataset 2

|       |   | y  |   |
|-------|---|----|---|
|       |   | 0  | 1 |
| $x_1$ | 0 | 19 | 1 |
|       | 1 | 71 | 9 |

In dataset 1, separation occurred due to a combination of rare outcome and imbalanced exposure. Therefore, ML yields perfect leave-one-out CV predictions $\hat{\pi}_{(-i)} = 0$ for $x_i = 0$ and the individual out-of-sample prediction errors equal $D_{x_i=0} = 0$. These errors, however, increase when applying shrinkage as predicted probabilities get pulled towards the observed event rate. In addition, the errors increase for those few observations where $x_i = 1$ and $y_i = 1$ as $\hat{\pi}_{(-i)} = 0.1$ in the ML model and $\hat{\pi}_{(-i)} = 0.08$ in the fully shrunken model. While shrinkage reduces the errors where $x_i = 1$ and $y_i = 0$ and although these observations are prevailing the differences in errors estimates are small regardless of whether shrinkage is applied or not (Figure 1). Therefore, the tuning procedure $D$ favors the smallest of the pre-specified range of $\lambda$ values, in our example $\lambda^* = 10^{(-6)}$. In this case, setting $\lambda^*$ to a value approaching zero yields meaningless coefficient estimate for $\beta_1$, resulting from an inability of numerical algorithm to converge [1, 16, 23]. In contrast, FC and RR with an informative prior (IP), assuming the 95% odds ratio interval of a standardized covariate ranging from 0.25 to 4, yield interpretable coefficient estimates (Table 1).

Figure 1. Leave one-out cross-validated deviance $D$ (top) and the sum of the deviance components $D_i$ for distinct observations (bottom) in dataset 1 (left) and dataset 2 (right) with respect to the complexity parameter $\lambda$.

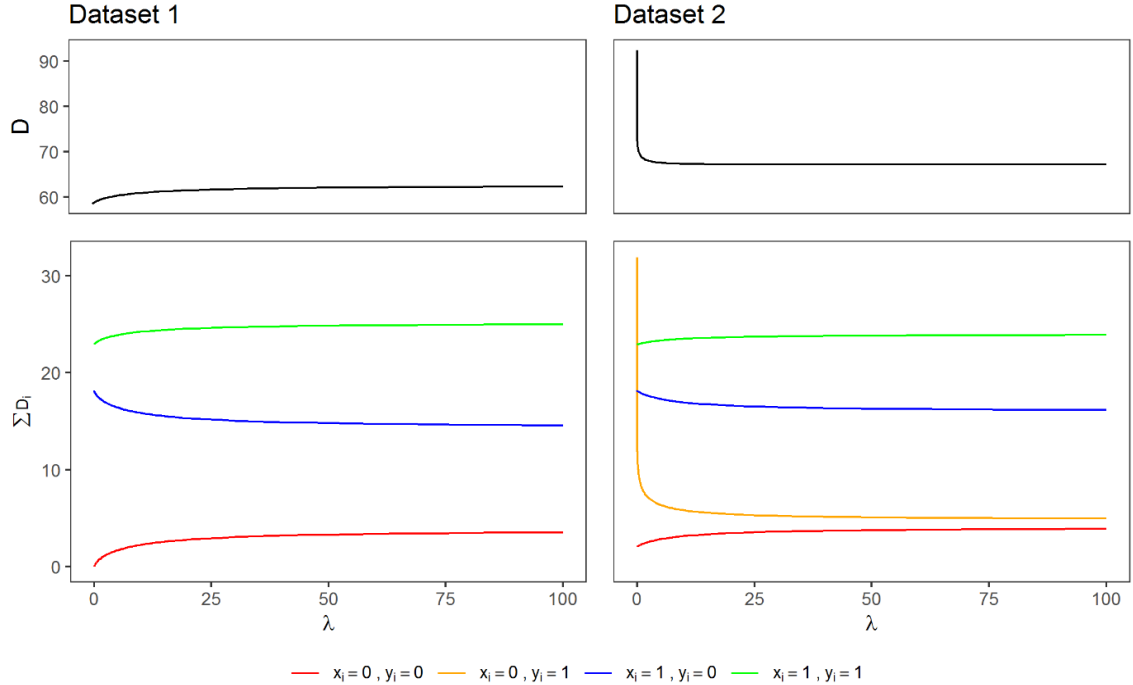

In dataset 2, we have one single observation $x_i = 0$ with an event for which ML falsely yields $\hat{\pi}_{(-i)} = 0$. While for all other observations the out-of-sample prediction errors $D_i$ do not change much if applying shrinkage (for some observations it gets slightly larger and for the others slightly smaller), the error for this single observation reduces considerably with increasing shrinkage (Figure 1). This results in $\lambda^*$ that equals the largest of the pre-specified range of values, in our example $\lambda^* = 100$. Obviously, this overshrinks the coefficients as compared to FC and IP (Table 1).

| Method | FC | RR | |
|---|---|---|---|
|  |  | $D$ | IP |
| Dataset 1 $\hat{\beta}_1$ | 1.7 | / | 1.54 |
| Dataset 2 $\hat{\beta}_1$ | 0.55 | 0.06 | 0.65 |

Table 1. Coefficients estimated by Firth's correction (FC) and ridge regression (RR) where complexity parameter is either tuned by leave one-out cross-validated deviance $D$ or set according to some informative prior (IP).

Repeating the data-generating process used to generate the two datasets 1000-times in which a binary covariate $X$ was sampled with $\mathrm{E}(X) = 0.8$ and the binary outcome $y_i$ was drawn from a Bernoulli distribution with true event probability $(1 + \exp(-(-3.05 + x_i)))^{-1}$ and each time tuning the value of $\lambda^*$ by $D$ results in a choice of $\lambda^*$ that often equals either the smallest or the largest value of the pre-specified range of $\lambda$ values (Figure 2). This arbitrary choice yields on average large variability of tuned $\lambda^*$ values and consequently, very unstable coefficients with large expected MSE. A great amount of large MSE of coefficients is due to separation that leads to optimized $\lambda^*$ values close to zero and thus prevents the iterative algorithm from converging. It is reasonable to assume that the instability in optimized complexity parameter values will also leave its consequences in predictions, translating into calibration slopes of large variability (models that strongly underfit or overfit). In contrast, in FC and IP no tuning is required and plausible estimates of $\beta_1$ are produced over 1000 datasets (Figure 2).

Figure 2. Distribution of complexity parameter values $\lambda^*$ tuned by leave one-out cross-validated deviance $D$ (left) and deviations of coefficients estimated by Firth's correction (FC) and ridge regression based on an informative prior (IP) from the true coefficient (right) together with bias (vertical red lines) over 1000 generated datasets. The latter results are not shown for $D$ since optimized $\lambda^*$ equaled $10^{(-6)}$ in 14.3%, yielding meaningless coefficients in these situations.

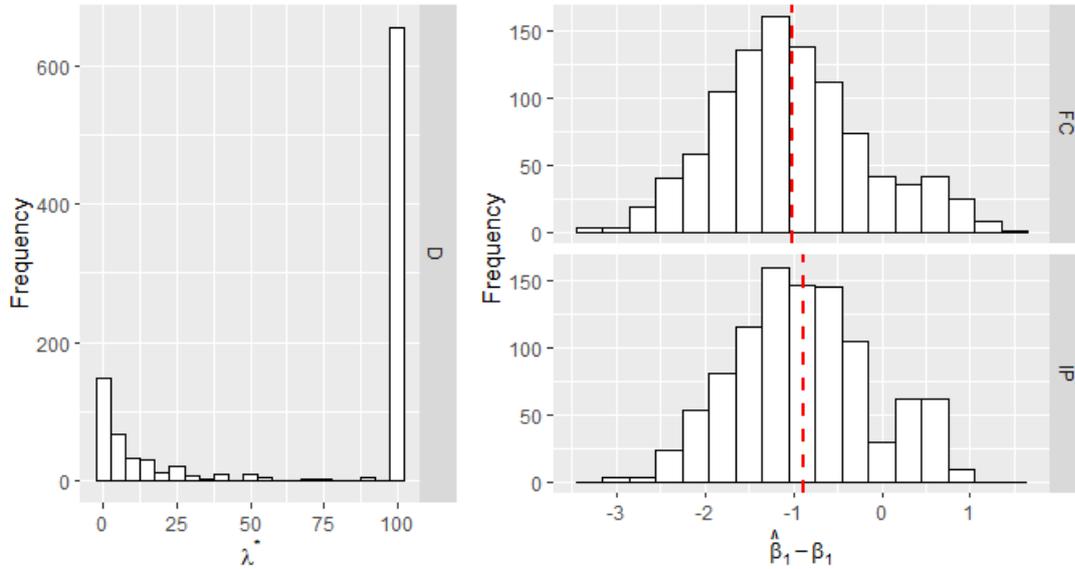

## 4 Simulation study

### 4.1 Design

We describe the simulation study design following recommendations by Morris [24].

#### 4.1.1 Aims

Our aim was to systematically investigate the performance of RR in terms of effect estimation and prediction in low-dimensional sparse data settings where the complexity parameter $\lambda$ was determined using different approaches and to compare it to FC.

#### 4.1.2 Data-generating mechanisms

To capture a plausible biomedical context, we considered a data generation scheme similar to the one described in Binder et al. [25]. Covariates $X_1, \ldots X_{15}$ were of mixed types obtained by applying certain transformations to variables $Z_1, \ldots, Z_{15}$ sampled from a standard multivariate normal distribution with correlation matrix $\mathbf{\Sigma}$ (Table 2). In particular, $X_1, \ldots, X_4$ were binary, $X_5$ and $X_6$ ordinal with three levels and $X_7, \ldots, X_{15}$ continuous. The continuous variables were truncated at the third quartile plus five times the interquartile distance of the corresponding distribution. The values of the binary outcome $y_i$ were sampled from Bernoulli distributions with event probabilities $(1 + \exp(-\beta_0 - a * (\beta_1 x_{i1} + \cdots + \beta_K x_{iK})))^{-1}$, where $i = 1, \ldots, N$, $N \in \{100, 250, 500\}$, $K \in \{2, 5, 10\}$, effect multiplier $a \in \{0.5, 1\}$ for moderate and strong effects, respectively, and true regression coefficients $\beta_1, \ldots \beta_K$ defined as follows: $\beta_1 = 2.08, \beta_2 = 1.39, \beta_3 = \beta_4 = 0.69, \beta_5 = \beta_6 = 0.35$ and the effects for $\beta_7, \ldots, \beta_{10}$ were chosen such that the log odds ratio between the first and the fifth sextile of the corresponding distribution was 0.69 (Table 2). In addition to these scenarios including real predictors only, we also considered the inclusion of noise covariates $X_{11}, \ldots X_{15}$ that had no effect on the outcome. An intercept $\beta_0$ was determined for each simulation scenario such that the desired marginal event rate $E(Y) \in \{0.1, 0.25\}$ was approximately obtained. Combining the simulation parameters $N$ (sample size), $K$ (number of true predictors), $a$ (effect multiplier), $E(Y)$ (marginal event rate) and noise (absent/present) in a full factorial design resulted in 72 possible scenarios. We simulated 1000 datasets for each scenario.

| $z_{ik}$ | Correlation of $z_{ik}$ | Type | $x_{ik}$ | $E(x_{ik})$ | $\beta_k$ |
|---|---|---|---|---|---|
| $z_{i1}$ | $z_{i2}(0.5), z_{i3}(0.5), z_{i7}(0.5), z_{i14}(0.5)$ | binary | $x_{i1} = I(z_{i1} < 0.84)$ | 0.80 | 2.08 |
| $z_{i2}$ | $z_{i1}(0.5), z_{i14}(0.3)$ | binary | $x_{i2} = I(z_{i2} < -0.35)$ | 0.36 | 1.39 |
| $z_{i3}$ | $z_{i1}(0.5), z_{i4}(-0.5), z_{i5}(-0.3)$ | binary | $x_{i3} = I(z_{i3} < 0)$ | 0.50 | 0.69 |
| $z_{i4}$ | $z_{i3}(-0.5), z_{i5}(0.5), z_{i7}(0.3), z_{i8}(0.5), z_{i9}(0.3), z_{i14}(0.5)$ | binary | $x_{i4} = I(z_{i4} < 0)$ | 0.50 | 0.69 |
| $z_{i5}$ | $z_{i3}(-0.3), z_{i4}(0.5), z_{i8}(0.3), z_{i9}(0.3)$ | ordinal | $x_{i5} = I(z_{i5} \geq -1.2) + I(z_{i5} \geq 0.75)$ | 1.11 | 0.35 |
| $z_{i6}$ | $z_{i7}(-0.3), z_{i8}(0.3), z_{i11}(-0.5)$ | ordinal | $x_{i6} = I(z_{i6} \geq 0.5) + I(z_{i6} \geq 1.5)$ | 0.38 | 0.35 |
| $z_{i7}$ | $z_{i1}(0.5), z_{i4}(0.3), z_{i6}(-0.3)$ | continuous | $x_{i7} = [10 z_{i7} + 55]$ | 54.5 | 0.036 |
| $z_{i8}$ | $z_{i4}(0.5), z_{i5}(0.3), z_{i6}(0.3), z_{i9}(0.5), z_{i12}(-0.3), z_{i14}(0.5)$ | continuous | $x_{i8} = [\max(0, 100 \exp(z_{i8}) - 20)]$ | 146 | 0.003 |
| $z_{i9}$ | $z_{i4}(0.3), z_{i5}(0.3), z_{i8}(0.5), z_{i14}(0.3)$ | continuous | $x_{i9} = [\max(0, 80 \exp(z_{i9}) - 20)]$ | 112 | 0.004 |
| $z_{i10}$ | - | continuous | $x_{i10} = [10 z_{i10} + 55]$ | 54.5 | 0.36 |
| $z_{i11}$ | $z_{i6}(-0.5), z_{i12}(0.3), z_{i15}(0.5)$ | continuous | $x_{i11} = \exp(0.4 z_{i11} + 3)$ | 21.8 | 0 |
| $z_{i12}$ | $z_{i8}(-0.3), z_{i11}(0.3), z_{i15}(0.5)$ | continuous | $x_{i12} = \exp(0.5 z_{i12} + 1.5)$ | 5.1 | 0 |
| $z_{i13}$ | - | continuous | $x_{i13} = 0.01 * [100(z_{i13} + 4)^2]$ | 17 | 0 |
| $z_{i14}$ | $z_{i1}(0.5), z_{i2}(0.3), z_{i4}(0.5), z_{i8}(0.5), z_{i9}(0.3)$ | continuous | $x_{i14} = [10 z_{i10} + 55]$ | 54.5 | 0 |
| $z_{i15}$ | $z_{i11}(0.5), z_{i12}(0.5)$ | continuous | $x_{i15} = [10 z_{i10} + 55]$ | 54.5 | 0 |

Table 2. Covariate structure and effect sizes applied in the simulation study based on Binder et al. [25]. [·] denotes removal of the non-integer part of the argument and $I$ is the indicator function.

4.1.3 Methods

We analyzed each simulated dataset by fitting RR and FC models. To obtain predictions based on FC we applied FLIC as suggested by Puhr et al. [7] RR models were computed after standardizing covariates to zero mean and unit variance and the complexity parameter $\lambda^*$ was selected from a fixed sequence of 200 log-linearly equidistant values ranging from $10^{(-6)}$ to 100 by using the following procedures:

- $D$;
- $GCV$;
- $CE$ where the cut-off $c$ was set to the observed event rate $\frac{1}{N} \sum_i y_i$. As $CE$ is discrete in nature and has no unique optimum in $\lambda$, in our study $\lambda^*$ was the largest $\lambda$ minimizing $CE$;
- $D$ by 10-fold CV with 50 repetitions (RCV) where $\lambda^*$ was chosen as the $\theta$-th quantile of the obtained values with $\theta \in \{0.5, 0.95\}$ [2, 21];

- AIC;
- restricting the standardized coefficients by informative (IP, $\lambda = 2$) and weakly informative prior assumptions (WP, $\lambda = 1/2$). In the simulations the degree of shrinkage was the same for all the covariates.

As a benchmark we defined two oracle models, determined by an amount of shrinkage ideal with respect to estimation of $\beta_1$ (explanation oracle, OEX) and to predictions (prediction oracle, OP). For OEX $\lambda^*$ was chosen such that $(\hat{\beta}_1 - \beta_1)^2$ (or equivalently $|\hat{\beta}_1 - \beta_1|$), where $\hat{\beta}_1$ is the RR estimate of $\beta_1$, was minimized; for OP, $\lambda^*$ was the one minimizing $\sum_i (\hat{\pi}_i - \pi_i)^2$, where $\hat{\pi}_i$ is the estimate of the $i$-th probability of $\pi_i$. All RR models were fitted by data augmentation [13]. Briefly, two artificial data records were added for each covariate; the values for this covariate were set to $1/s$ and to zero for other covariates, where $s = 10$ was a rescaling factor improving the approximation. ML estimation on this augmented dataset was then performed, specifying weights that equaled 1 for the original observations and $2s^2\lambda$ for the pseudo-observations. We used the libraries brglm2 [26] for detecting separation, penalized [27] for performing CV, logistf [28] for estimation in R version 4.0.2 [29].

### 4.1.4 Estimands

The true regression coefficients $\beta_1$ and the vector of true event probabilities $\pi$ were the estimands in our study.

### 4.1.5 Performance measures

We evaluated the root mean squared errors (RMSE) of coefficients $\left(\frac{1}{1000}\sum_{s=1}^{1000}(\hat{\beta}_{k,s} - \beta_k)^2\right)^{1/2}$, where $\hat{\beta}_{k,s}$, $k = 1$, $s \in \{1, \dots, 1000\}$ is the estimate of $\beta_k$ in the $s$-th simulated dataset) and predictions $\left(\frac{1}{1000N}\sum_{s=1}^{1000}\sum_{i=1}^{N}(\hat{\pi}_{i,s} - \pi_{i,s})^2\right)^{1/2}$, where $\hat{\pi}_{i,s}$ and $\pi_{i,s}$ are the estimated and true event probability for the $i$-th observation in the $s$-th simulated dataset). We also evaluated c-statistics, estimated with respect to newly generated outcome, and calibration slopes evaluated on a validation datasets generated from the same population with a sample size $N = 10,000$. To combine bias and variability of calibration slopes, we calculated root mean squared distances (RMSD, $\left(\frac{1}{1000}\sum_{s=1}^{1000}d_s^2\right)^{1/2}$), where the $s$-th distance was defined as $d_s = \log(1) - \log(\text{slope}_s)$, as suggested by Van Calster et al. [10]. To avoid issues with negative slopes that were rarely obtained by the methods we winsorized them at 0.01 for the calculation of RMSD. In addition, we assessed the spearman correlation coefficients between calibration slopes and tuned complexity parameters $\lambda^*$ as well as the RMSD of calibration slopes achieved by the methods and the variability of tuned complexity parameters $\lambda^*$, expressed by median absolute deviation (MAD), over all simulated scenarios.

## 4.2 Results

Among 72 simulated scenarios the prevalence of separation was ranging from zero in scenarios with moderate effects, large sample sizes and $\text{E}(Y) = 0.25$ to at most 85% in a scenario with large effects, $N = 100$, $K = 5$ and $\text{E}(Y) = 0.1$ (Table S1 in Additional file 1).

Figure 3. Scatter plots showing values of $\lambda^*$ obtained by optimizing different tuning criteria versus 'optimal' $\lambda^*$ as achieved by explanation (OEX) and prediction (OP) oracle over 1000 generated datasets in scenarios with the expected value of $Y$, $\mathrm{E}(Y) = 0.1$, the number of predictors $K = 5$, noise absent or present, the sample size of $N \in \{100, 250, 500\}$ considering A) moderate ($a = 0.5$) and B) strong ($a = 1$) predictors. Values of $\lambda^*$ were optimized by using different tuning criteria: $D$, deviance; $GCV$, generalized cross-validation; $CE$, classification error; RCV50, repeated 10-fold cross-validated deviance with $\theta = 0.5$; RCV95, repeated 10-fold cross-validated deviance with $\theta = 0.95$; AIC, Akaike's information criterion.

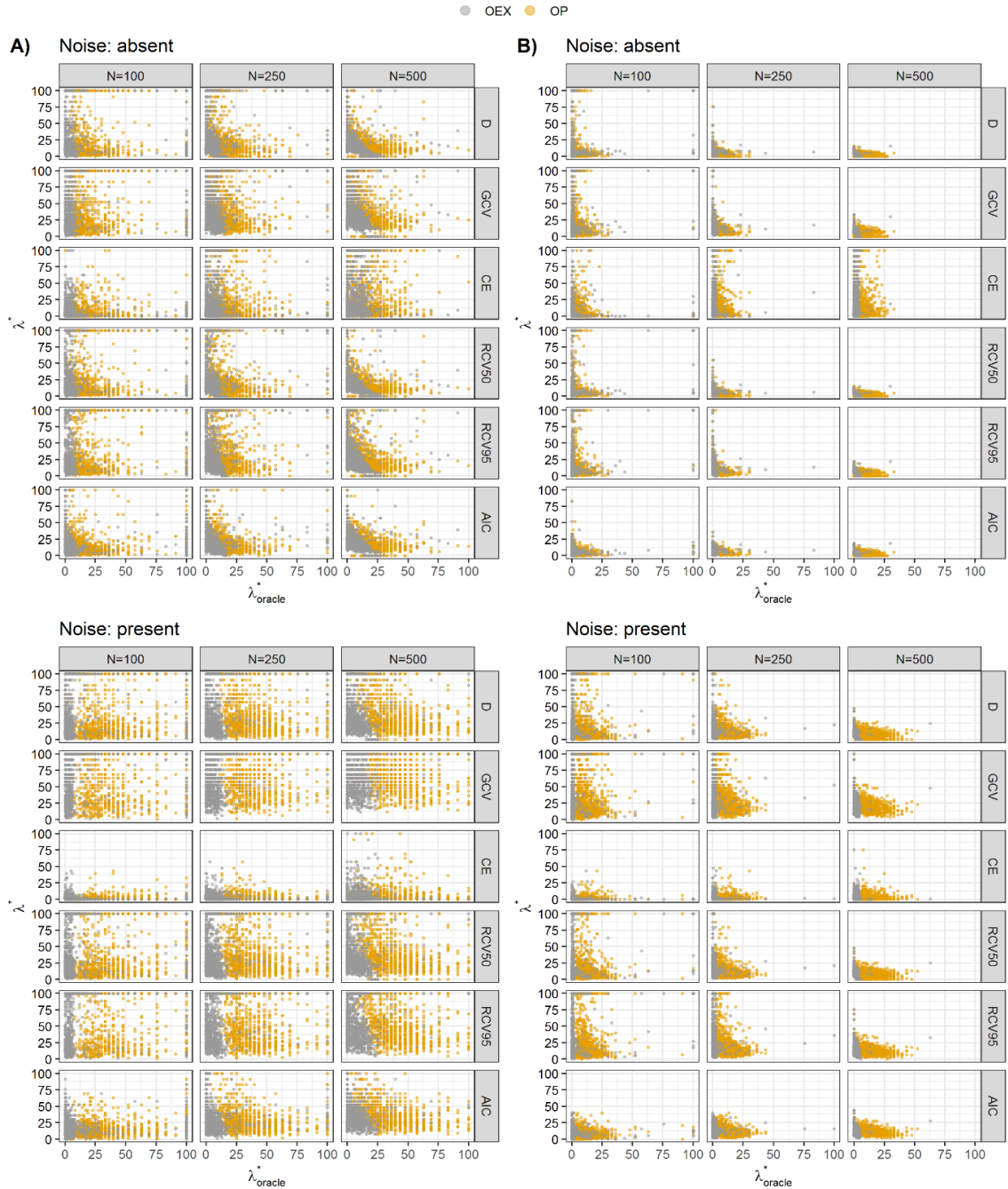

First, we describe the distribution of $\lambda^*$ values obtained by optimizing different tuning criteria over 1000 simulation runs and their correlations with 'optimal' $\lambda^*$ as achieved by OEX and OP, respectively (Figure 3). For brevity, Figure 3 focuses on scenarios with $\mathrm{E}(Y) = 0.1$ and $K = 5$ only. Tuning procedures often led to large variability of selected $\lambda^*$, which was especially apparent in moderate effects scenarios. Generally, the variability was smaller when the true effects were strong, with larger $N$ and $K$, i.e. the number of predictors associated with the outcome, and with more balanced outcomes. With moderate effects the methods tended to overshrink, often producing very wide distributions of optimized $\lambda^*$ values. The smallest variability of optimized values over all scenarios calculated as MAD was obtained by the AIC, followed by $CE$ and RCV50. Quite some variability of $\lambda^*$ was also obtained by OP, however, the correlations between 'optimal' $\lambda^*$ (of both OP and OEX) and $\lambda^*$ obtained by optimizing different tuning criteria were mostly negative. OEX resulted in less variability of $\lambda^*$ and less shrinkage than OP. In IP the pre-specified $\lambda^*$ was in median very close to the one obtained by OEX.

Figure 4. Nested loop plot of root mean squared error (RMSE) of $\hat{\beta}_1$ by the expected value of $Y$, $\mathrm{E}(Y) \in \{0.1, 0.25\}$, the number of predictors $K \in \{2, 5, 10\}$, noise absent or present (full and dashed lines), the sample size $N \in \{100, 250, 500\}$ and the size of true coefficients $\beta_1 \in \{1.04, 2.08\}$ for all simulated scenarios. Due to poor performance some results lie outside the plot range. OEX, explanation oracle; $D$, deviance; $GCV$, generalized cross-validation; $CE$, classification error; RCV50, repeated 10-fold cross-validated deviance with $\theta = 0.5$; RCV95, repeated 10-fold cross-validated deviance with $\theta = 0.95$; AIC, Akaike's information criterion; IP, shrinkage based on informative priors; WP, shrinkage based on weakly informative priors; FC, Firth's correction. Results regarding RMSE of $\beta_2$ are contained in Table S3 and S4.

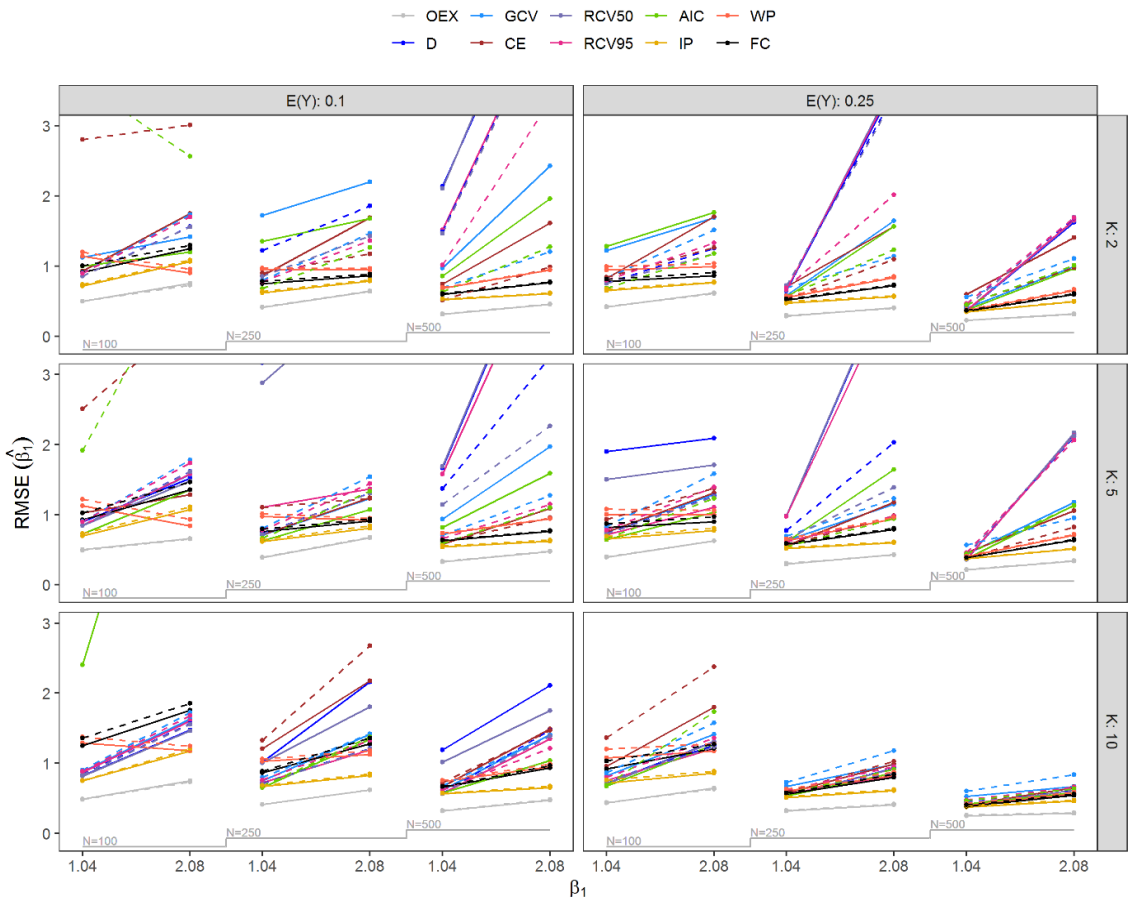

### 4.2.1 Accuracy of coefficients

Figure 4 shows the RMSE of $\beta_1$ across all simulated scenarios and models with and without noise by means of nested loop plots [30, 31]. More detailed results are contained in Table S1 of Additional file 1. As expected the best performance across all simulated scenarios was achieved by OEX. Generally, the performance of other tuned RR approaches was extremely variable and unreliable and due to separation the methods yielded coefficients with extremely large RMSE. Interestingly, the RMSE of $\beta_1$ did not always decrease with increasing sample size and noise did not necessarily worsen the performance of those methods. In contrast, the methods where tuning was not required showed stable performance across all simulated scenarios. While the performance of FC was satisfactory in almost all scenarios, suffering from RMSE larger than one in scenarios with the expected event rate $\mathrm{E}(Y) = 0.1$ and sample size $N = 100$ only, it was clearly outperformed by IP that produced small RMSE of $\beta_1$ across all scenarios. Although WP generally resulted in worse performance than FC, it was less sensitive to very sparse data situations in which the performance of FC was poor.

### 4.2.2 Accuracy of predictions

Results regarding the RMSE of predicted probabilities for $\mathrm{E}(Y) = 0.1$ are shown in Figure 5. While the performance of the methods was similar in scenarios with larger sample sizes and no noise, the differences between them became apparent in scenarios with $N = 100$ and especially when including noise. Noise considerably worsened the performance of the methods. The least affected by noise were the methods based on CV (apart from $CE$) that generally yielded the best performance. However, with no noise and strong effects IP always outperformed the other methods (apart from OP). The performance of FC and WP was consistently somewhat worse than the one of IP. The performance of AIC was similar to that of the CV-based methods if there was no noise, however, when $N = 100$ it appeared sensitive to noise.

Figure 5. Nested loop plot of root mean squared error (RMSE) of predictions multiplied by the square root of sample size $N$ by the number of predictors $K \in \{2, 5, 10\}$, noise absent or present, the sample size $N \in \{100, 250, 500\}$ and the effect multiplier $a \in \{0.5, 1\}$ for all simulated scenarios with expected value of $Y$, $\mathrm{E}(Y) = 0.1$. See Table S6 for results on scenarios with $\mathrm{E}(Y) = 0.25$. OP, prediction oracle; $D$, deviance; $GCV$, generalized cross-validation; $CE$, classification error; RCV50, repeated 10-fold cross-validated deviance with $\theta = 0.5$; RCV95, repeated 10-fold cross-validated deviance with $\theta = 0.95$; AIC, Akaike's information criterion; IP, shrinkage based on informative priors; WP, shrinkage based on weakly informative priors; FLIC, Firth's logistic regression with intercept-correction.

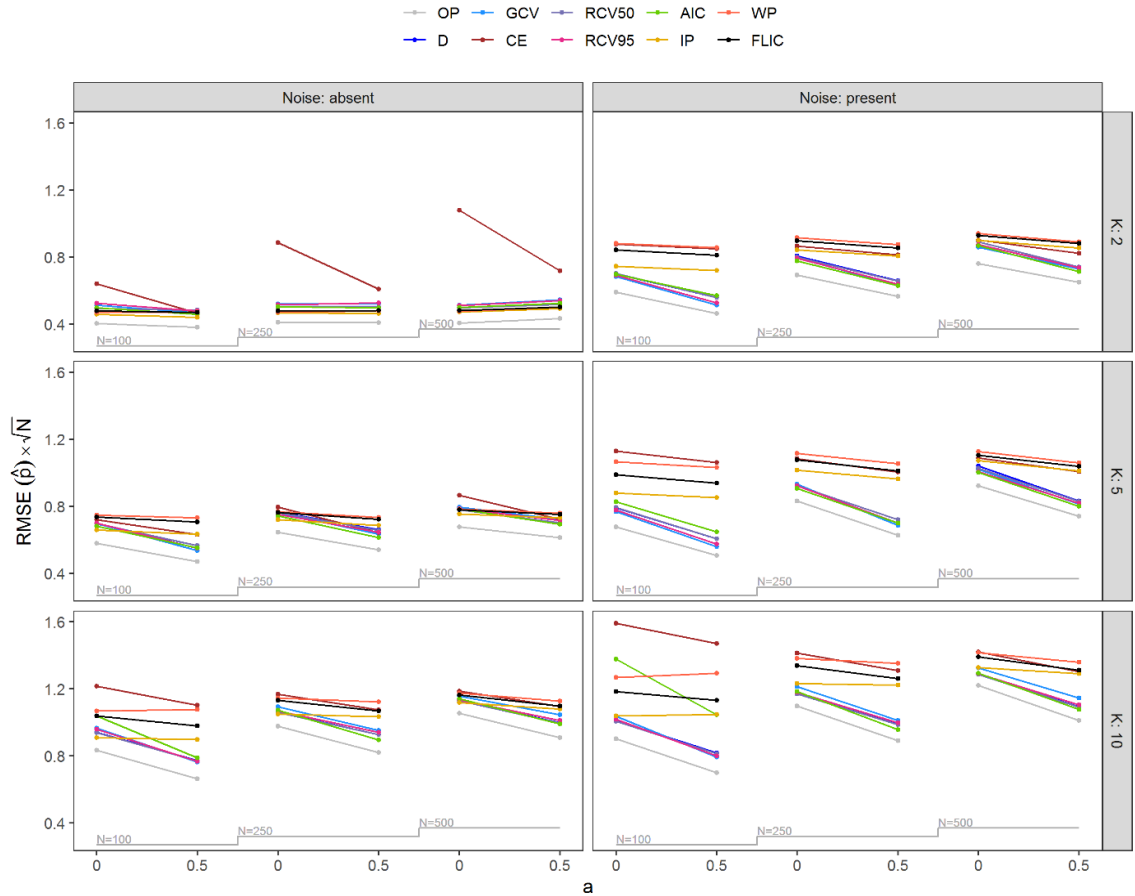

Calibration slopes are presented by means of boxplots for scenarios with $\mathrm{E}(Y) = 0.1$ and $K = 5$ (Figure 6). For clarity of presentation, datasets where calibration slopes were larger than 5 are not shown. We also included mean calibration slopes together with error bars indicating standard deviations to the plots, however these measures were heavily influenced by a large number of outliers, especially in scenarios with small sample sizes. IP, WP and FC yielded similar performance with small variability over simulation runs and were generally close to the slope of 1 in strong effects scenarios but were overfitting in moderate effects scenarios. The addition of noise increased overfitting of these methods with FC being the least sensitive to noise. Tuning procedures (specifically, $D$, RCV50 and AIC) yielded calibration slopes that were in median relatively close to 1, however they suffered from large variability. While this variability decreased with $N$, there was still a considerable amount of outliers produced by these methods even with $N = 500$, if the outcomes were very unbalanced and effects moderate only. Among tuning procedures AIC achieved the smallest variability of calibration slopes. With respect to the RMSD of the logarithm of calibration slopes (Table S9, S10 in Additional file 1) OP overall achieved the best performance, followed by IP if no noise was included or AIC in case of noise. Interestingly, noise did not necessarily increase the RMSD of tuning procedures and they appeared to be less sensitive to it as compared to the methods where shrinkage was pre-specified. Calibration slopes were strongly positively correlated with optimized $\lambda^*$ values (Figure S1 in Additional file 1) and such were the correlations between the RMSD of calibration slopes and the variability of $\lambda^*$.

Figure 6. Boxplots showing distribution of calibration slopes over 1000 generated datasets in scenarios with the expected value of $Y$, $E(Y) = 0.1$, the number of predictors $K = 5$, noise absent or present, the sample size of $N \in \{100, 250, 500\}$ considering A) moderate ($a = 0.5$) and B) strong ($a = 1$) predictors. The whiskers extend no more than 1.5-times the interquartile range from the box. In addition, mean calibration slopes together with error bars indicating one standard deviation are shown in red. Further results from other scenarios are contained in Table S7 and S8. OP, prediction oracle; $D$, deviance; $GCV$, generalized cross-validation; $CE$, classification error; RCV50, repeated 10-fold cross-validated deviance with $\theta = 0.5$; RCV95, repeated 10-fold cross-validated deviance with $\theta = 0.95$; AIC, Akaike's information criterion; IP, shrinkage based on informative priors; WP, shrinkage based on weakly informative priors; FLIC, Firth's logistic regression with intercept-correction.

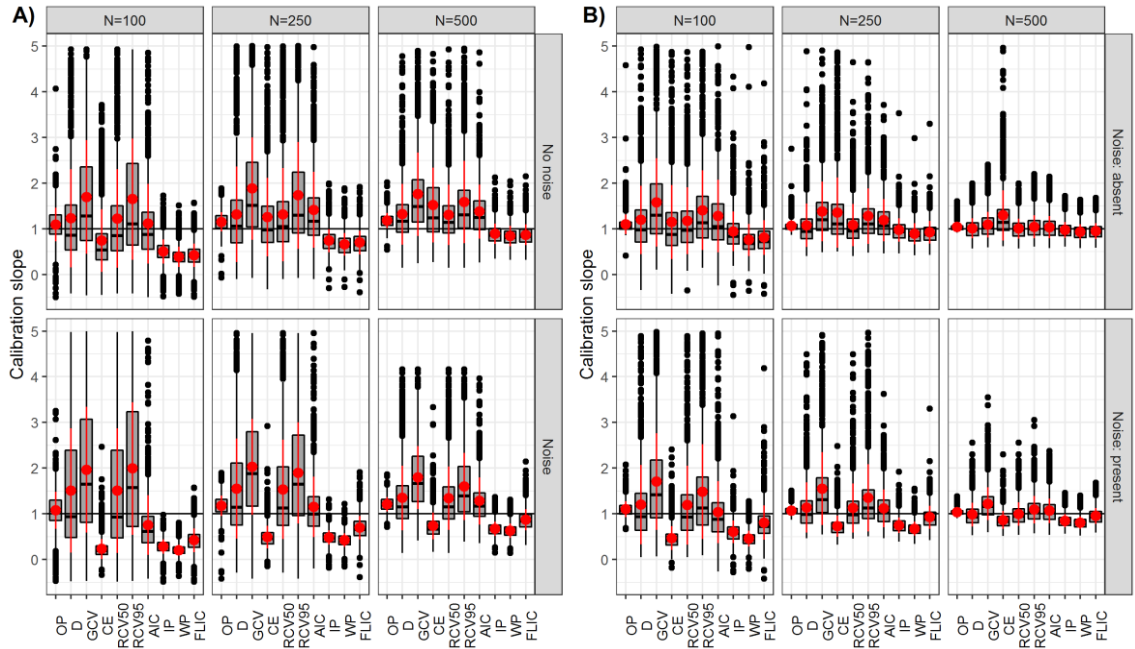

In terms of c-indices there was no considerable differences between methods (Table S11, S12 in Additional file 1).

## 5 Discussion

Numerous studies have showed that shrinkage is effective in preventing overfitting and may solve issues that arise in classical clinical settings with relatively large number of correlated covariates [2, 7, 8, 32]. Therefore, applying shrinkage has been recommended not only when developing prediction models but also when interest lies in coefficients with reduced MSE and inference is not required [7]. A recent study, however, noted that while calibration slopes obtained by shrinkage methods are on average close to 1 the variability of calibration slopes in small or sparse situations is large and therefore, improved performance in a single dataset cannot be guaranteed [10]. While this may not seem surprising, many researchers would still utilize tuned RR, expecting all problems arising from small or sparse datasets to be solved. Therefore, in this paper, we have elaborated this issue further by focusing on RR. We evaluated its performance in a low-dimensional setting and compared it to FC by means of simulation study. The amount of shrinkage in RR was determined using different tuning procedures and prior assumptions, respectively. We were interested in both the accuracy of coefficient and prediction estimates.

With respect to large variability of calibration slopes, the results of our study confirm the findings of Van Calster et al. [10]. Furthermore, we observed that the RMSD of the logarithm of calibration slopes was strongly correlated with the variability of optimized complexity parameters $\lambda^*$. By means of an illustrative example we demonstrated that tuning procedures might fail to approximate the U-shaped curve arising from the bias-variance trade-off and result in completely arbitrary choice of $\lambda^*$ that simply equals the smallest or the largest $\lambda$ of the pre-specified sequence of values. In the simulation study we then observed that substantial variability of $\lambda^*$ must even be expected from a prediction oracle that 'knows' the true event probabilities and can determine the optimal values of $\lambda^*$. However, the values of $\lambda^*$ obtained by tuning procedures were negatively correlated with their 'optimal counterparts', determined by explanation or prediction oracle. On one hand this could be explained by separation that often makes tuning procedures result in an optimized value of $\lambda^*$ close to zero and which causes nonconvergence of the iterative estimation process to find coefficient estimates [11]. Thus, separation is undoubtedly a non-negligible factor when the accuracy of coefficients is of concern. In addition, if the amount of shrinkage is too small, rareness of events together with numerical instability or inaccuracy may make predictions numerically indistinguishable from zero or one by software packages [15]. While some may argue that this is not problematic when interest lies in predictions, we observed that the prediction oracle favored larger values of $\lambda^*$ than the explanation oracle, suggesting that in typical clinical studies seemingly perfect predictions should not be accepted as they could indicate overfitting and increased variability of calibration slopes. On the other hand, as illustrated in our example, if only a few observations prevent the data set from separation, a large value of $\lambda^*$ is needed to avoid very large out-of-sample prediction errors for the crucial, separation-preventing observations, while in fact the performance of ML in such datasets is already satisfactory [15]. In our simulation study we observed that the performance of tuned RR in terms of calibrations slopes was more stable and the variability of $\lambda^*$ was lower in scenarios with larger sample sizes, more balanced outcomes, stronger effects and often even with a larger number of true predictors. The presence of noise generally increased the variability of $\lambda^*$ thus the RMSD of calibration slopes. This suggests that simply breaking down the problem to a measure such as the events-per-variable ratio is unsatisfactory as not only the number of covariates but also their relations to the outcome are decisive here [8].

Our results show that optimization of the complexity parameter in RR is difficult in datasets where sampling variability is large and sampling artefacts, e.g. separation, are likely to occur. In such datasets tuning procedures yield highly variable optimized values of complexity parameters that are negatively correlated with the optimal amount of shrinkage. In words of van Houwelingen [33], if $\beta$ is 'large' by random fluctuation tuning procedures tend to keep the model large instead of correcting for the 'large' $\beta$ by setting $\lambda^* > 0$; and vice versa. Van Calster et al. [10] instead suggested to apply intercept corrected FC [7] that provides only little shrinkage but results in low variability of calibration slopes. Another convenient property of FC, not shared by RR, is its invariance to linear transformations of the design matrix. Alternatively, the choice of the complexity parameter may be based on prior expectations about the magnitude of the underlying effects [13]. Pre-specifying the degree of shrinkage seems reasonable as it stabilizes $\lambda^*$, and appeared beneficial in our study in which we included such a Bayesian approach with zero-centered informative or weakly informative normal priors (IP and WP, respectively). Despite different motivation behind methods with fixed penalization strength, IP clearly outperformed tuned RR (and FC) with regard to frequentist criterion of RMSE of coefficients. In contrast to tuned RR, where obtaining valid inference is hard due to bias introduced in the coefficients and additional variability that comes along with tuning $\lambda$ (which possibly leads to less bias and more variance) [12, 34], like in FC valid 95% posterior limits could be achieved easily by data augmentation, using any statistical software that enables ML fitting and weighting of observations [13]. Moreover, in scenarios with no noise included in the model IP yielded small RMSE of predictions and small RMSD of calibration slopes. Although one should usually devote additional work to specify prior distributions, we straightforwardly followed the outline of Greenland et al. [13] in defining our priors, assuming that the true effects are not too extreme. IP therefore performed extremely well in all scenarios with strong effects but it would be reasonable to adapt the prior for scenarios with moderate effects only. However, if one is in doubt about how to determine the range of the prior interval, weaker penalties are preferred. More guidance on how to specify prior distributions can be found in the paper by Greenland et al. [1].

Our study showed that all the approaches and even the oracle were sensitive to noise, particularly with regard to predictions. Worse performance in terms of calibration slopes after the addition of noise was especially notable in methods with a fixed degree of shrinkage. Other shrinkage methods that also perform variable selection, e.g. lasso, may be used to filter out noise from the models instead, however in a classical clinical context lasso may be too restrictive and result in too sparse models. Besides we suspect that similar issues with respect to tuning will also appear with the lasso or any other tuned penalized regression method [10]. Therefore, noise should be removed by using subject domain knowledge prior to fitting. On the other hand, we should be liberal in the inclusion of covariates where the associations with the outcome are less clear in order to reduce bias of coefficients but also to assure reproducibility of prediction modelling studies by preventing to miss true predictors [1, 35].

Summarizing, while tuning has the potential to reduce the MSE of its estimates as demonstrated by the oracles, applying tuned RR in small or sparse datasets is problematic as tuned $\lambda^*$ values are highly variable and in addition negatively correlated with optimal values, yielding unstable coefficients and predictions. To overcome the trouble, we recommend to determine the degree of shrinkage according to some meaningful prior assumptions about true effects, since we found that such an approach has the potential to reduce bias and stabilize the estimates. We showed that with a predetermined complexity parameter accurate coefficients and predictions can be obtained even in non-ideal settings which are typical e.g. in the context of rare outcomes or sparse predictors. Moreover, it also enables valid Bayesian inference [1, 13].

After finishing the first draft of this paper, a paper by Riley et al. [36] was published which came to partly similar conclusions regarding the instability of tuned penalized regression with small samples as our work. By providing a high-level overview of the shrinkage methods that paper targets at a wider audience, not explaining technical details regarding tuning. For clarity, their focus is on tuning the CV deviance only and other tuning criteria commonly used in practice were excluded. Moreover, their findings are based on a simplified, exemplary simulation setup with continuous predictors only. In contrast, our simulation study is designed in a comprehensive way, allowing a fair comparison between the methods by capturing a plausible biomedical context [37]. We also considered FC, IP and WP as alternative solutions with pre-specified degree of shrinkage. By including oracle models in the simulation study we demonstrated what the best possible performance of RR could be if truth was known. We showed that tuned complexity parameter values do not only suffer from large variability as demonstrated by Riley et al. [36] but are in addition negatively correlated with the optimal amount of shrinkage, yielding questionable performance of tuned RR in small and sparse datasets.

**Abbreviations:** Akaike's information criterium, AIC; classification error, $CE$; cross-validation, CV; deviance, $D$; , Firth's correction, FC; Firth's logistic regression with intercept-correction, FLIC; generalized cross-validation, $GCV$; ridge regression based on informative prior assumptions, IP; median absolute deviation, MAD; maximum likelihood, ML; mean squared error, MSE; explanation oracle, OEX; prediction oracle, OP; repeated 10-fold cross-validated deviance, RCV; root mean squared distance, RMSD; root mean squared error, RMSE; ridge logistic regression, RR; ridge regression based on weakly informative prior assumptions, WP.

**Funding:** This research was funded by the Austrian Science Fund (FWF), grant number I 2276-N33.

**Supplementary Information:** Additional file 1 contains detailed simulation results.

## 6 References

1. Greenland S, Mansournia MA, Altman DG: **Sparse data bias: a problem hiding in plain sight**. *BMJ* 2016, **352**:i1981.

2. Pavlou M, Ambler G, Seaman S, De Iorio M, Omar RZ: **Review and evaluation of penalised regression methods for risk prediction in low-dimensional data with few events**. *Statistics in Medicine* 2016, **35**(7):1159-1177.

3. Le Cessie S, Van Houwelingen JC: **Ridge Estimators in Logistic Regression**. *Journal of the Royal Statistical Society Series C (Applied Statistics)* 1992, **41**(1):191-201.

4. Hastie T, Tibshirani R, Friedman JH: **The Elements of Statistical Learning: Data Mining, Inference, and Prediction**: Springer; 2009.

5. Belkin M, Hsu D, Ma S, Mandal S: **Reconciling modern machine-learning practice and the classical bias–variance trade-off**. *Proceedings of the National Academy of Sciences* 2019, **116**(32):15849-15854.

6. Harrell FE, jrl FEH: **Regression Modeling Strategies: With Applications to Linear Models, Logistic Regression, and Survival Analysis**: Springer; 2001.

7. Puhr R, Heinze G, Nold M, Lusa L, Geroldinger A: **Firth's logistic regression with rare events: accurate effect estimates and predictions?** *Statistics in Medicine* 2017, **36**(14):2302-2317.

8. van Smeden M, Moons KG, de Groot JA, Collins GS, Altman DG, Eijkemans MJ, Reitsma JB: **Sample size for binary logistic prediction models: Beyond events per variable criteria**. *Statistical Methods in Medical Research* 2019, **28**(8):2455-2474.

9. Blagus R, Lusa L: **Class prediction for high-dimensional class-imbalanced data**. *BMC Bioinformatics* 2010, **11**:523.

10. Van Calster B, van Smeden M, De Cock B, Steyerberg EW: **Regression shrinkage methods for clinical prediction models do not guarantee improved performance: Simulation study**. *Statistical Methods in Medical Research* 2020, **29**(11):3166-3178.

11. Šinkovec H, Geroldinger A, Heinze G, Blagus R: **Tuning in ridge logistic regression to solve separation**. *arXiv: 201114865* 2020.

12. Blagus R, Goeman JJ: **Mean squared error of ridge estimators in logistic regression**. *Statistica Neerlandica* 2020, **74**(2):159-191.

13. Sullivan SG, Greenland S: **Bayesian regression in SAS software**. *Int J Epidemiol* 2013, **42**(1):308-317.

14. FIRTH D: **Bias reduction of maximum likelihood estimates**. *Biometrika* 1993, **80**(1):27-38.

15. Šinkovec H, Geroldinger A, Heinze G: **Bring More Data!—A Good Advice? Removing Separation in Logistic Regression by Increasing Sample Size.** *Int J Environ Res Public Health* 2019, **16**(23):4658.

16. Heinze G, Schemper M: **A solution to the problem of separation in logistic regression**. *Statistics in Medicine* 2002, **21**(16):2409-2419.

17. Agresti A: **Categorical Data Analysis**: Wiley; 2012.

18. Golub GH, Heath M, Wahba G: **Generalized Cross-Validation as a Method for Choosing a Good Ridge Parameter**. *Technometrics* 1979, **21**(2):215-223.

19. van wieringen WN: **Lecture notes on ridge regression**. *arXiv: 150909169* 2020.

20. Wood S: **Generalized Additive Models: An Introduction with R**: Taylor & Francis; 2006.

21. Roberts S, Nowak G: **Stabilizing the lasso against cross-validation variability**. *Computational Statistics & Data Analysis* 2014, **70**:198-211.

22. Akaike H: **A new look at the statistical model identification**. *IEEE Transactions on Automatic Control* 1974, **19**(6):716-723.

23. Mansournia MA, Geroldinger A, Greenland S, Heinze G: **Separation in Logistic Regression: Causes, Consequences, and Control**. *American Journal of Epidemiology* 2017, **187**(4):864-870.

24. Morris TP, White IR, Crowther MJ: **Using simulation studies to evaluate statistical methods**. *Statistics in Medicine* 2019, **38**(11):2074-2102.

25. Binder H, Sauerbrei W, Royston P: **Multivariable model-building with continuous covariates: 1. Performance measures and simulation design**. *Technical Report FDM-Preprint 105* 2011.

26. Kosmidis I: **brglm2: Bias Reduction in Generalized Linear Models**. 2020.

27. Goeman JJ, Meijer R, Chaturvedi N: **Penalized: L1 (lasso and fused lasso) and L2 (ridge) penalized estimation in GLMs and in the Cox model**. 2018.

28. Heinze G, Ploner M, Jiricka L: **logistf: Firth's Bias-Reduced Logistic Regression**. 2020.

29. Team RC: **R: A Language and Environment for Statistical Computing**. 2020.

30. Kammer M: **looplot: A package for creating nested loop plots**. 2020.

31. Rücker G, Schwarzer G: **Presenting simulation results in a nested loop plot**. *BMC Medical Research Methodology* 2014, **14**(1):129.

32. Steyerberg EW, Eijkemans MJC, Harrell Jr FE, Habbema JDF: **Prognostic modelling with logistic regression analysis: a comparison of selection and estimation methods in small data sets**. *Statistics in Medicine* 2000, **19**(8):1059-1079.

33. Van Houwelingen JC: **Shrinkage and Penalized Likelihood as Methods to Improve Predictive Accuracy**. *Statistica Neerlandica* 2001, **55**(1):17-34.

34. Heinze G, Wallisch C, Dunkler D: **Variable selection – A review and recommendations for the practicing statistician**. *Biometrical Journal* 2018, **60**(3):431-449.

35. Steyerberg EW, Van Calster B: **Redefining significance and reproducibility for medical research: A plea for higher P-value thresholds for diagnostic and prognostic models**. *European Journal of Clinical Investigation* 2020, **50**(5):e13229.

36. Riley RD, Snell KIE, Martin GP, Whittle R, Archer L, Sperrin M, Collins GS: **Penalisation and shrinkage methods produced unreliable clinical prediction models especially when sample size was small**. *Journal of Clinical Epidemiology* 2020.

37. Boulesteix A-L, Binder H, Abrahamowicz M, Sauerbrei W, Initiative ftSPotS: **On the necessity and design of studies comparing statistical methods**. *Biometrical Journal* 2018, **60**(1):216-218.

# To tune or not to tune, a case study of ridge logistic regression in small or sparse datasets

Hana Šinkovec, Georg Heinze, Rok Blagus, Angelika Geroldinger

**Additional file 1**

Table S1. Simulation results showing prevalence of separation (SP, %) and root mean squared errors of $\beta_1$ across simulation scenarios with marginal event rate $E(Y) = 0.10$ that differed by the number of predictors $K \in \{2, 5, 10\}$, the sample size $N \in \{100, 250, 500\}$, the effect multiplier $a \in \{0.5, 1\}$ and noise absent (0) or present (1). OEX, explanation oracle; *D*, deviance; *GCV*, generalized cross-validation; *CE*, classification error; RCV50, repeated 10-fold cross-validated deviance with $\theta = 0.5$; RCV95, repeated 10-fold cross-validated deviance with $\theta = 0.95$; AIC, Akaike's information criterion; IP, shrinkage based on informative priors; WP, shrinkage based on weakly informative priors; FC, Firth's correction.

| K | N | $\beta_1$ | a | Noise | SP | OEX | D | GCV | CE | RCV50 | RCV95 | AIC | IP | WP | FC |
|---|---|---|---|---|---|---|---|---|---|---|---|---|---|---|---|
| 2 | 100 | 2.08 | 1 | 0 | 79 | 0.75 | 6.77 | 1.42 | 1.75 | 7.02 | 3.53 | 1.2 | 1.06 | 0.9 | 1.25 |
| 2 | 100 | 2.08 | 1 | 1 | 79 | 0.73 | 1.56 | 1.73 | 3.01 | 1.56 | 1.7 | 2.57 | 1.08 | 0.96 | 1.3 |
| 2 | 250 | 2.08 | 1 | 0 | 54 | 0.64 | 7.18 | 2.2 | 1.69 | 7.28 | 5.64 | 1.68 | 0.79 | 0.94 | 0.86 |
| 2 | 250 | 2.08 | 1 | 1 | 54 | 0.65 | 1.86 | 1.46 | 1.18 | 1.43 | 1.37 | 1.27 | 0.8 | 0.97 | 0.89 |
| 2 | 500 | 2.08 | 1 | 0 | 33 | 0.45 | 5.58 | 2.43 | 1.61 | 5.73 | 4.79 | 1.96 | 0.61 | 0.94 | 0.77 |
| 2 | 500 | 2.08 | 1 | 1 | 33 | 0.46 | 4.73 | 1.21 | 0.99 | 4.7 | 3.35 | 1.28 | 0.62 | 0.95 | 0.78 |
| 5 | 100 | 2.08 | 1 | 0 | 85 | 0.66 | 1.52 | 1.58 | 1.29 | 1.48 | 1.58 | 1.36 | 1.07 | 0.84 | 1.36 |
| 5 | 100 | 2.08 | 1 | 1 | 85 | 0.65 | 1.62 | 1.78 | 3.81 | 1.61 | 1.74 | 4.71 | 1.11 | 0.93 | 1.47 |
| 5 | 250 | 2.08 | 1 | 0 | 63 | 0.67 | 4.78 | 1.23 | 1.24 | 4.22 | 1.37 | 1.07 | 0.81 | 0.92 | 0.91 |
| 5 | 250 | 2.08 | 1 | 1 | 63 | 0.67 | 1.31 | 1.54 | 1.23 | 1.31 | 1.45 | 1.33 | 0.84 | 0.95 | 0.94 |
| 5 | 500 | 2.08 | 1 | 0 | 41 | 0.47 | 5.29 | 1.97 | 1.09 | 5.33 | 4.96 | 1.59 | 0.62 | 0.94 | 0.76 |
| 5 | 500 | 2.08 | 1 | 1 | 41 | 0.48 | 3.25 | 1.27 | 0.96 | 2.26 | 1.15 | 1.1 | 0.64 | 0.95 | 0.77 |
| 10 | 100 | 2.08 | 1 | 0 | 72 | 0.75 | 1.47 | 1.63 | 7.69 | 1.46 | 1.6 | 7.41 | 1.17 | 1.18 | 1.75 |
| 10 | 100 | 2.08 | 1 | 1 | 72 | 0.73 | 1.57 | 1.72 | 20.33 | 1.56 | 1.68 | 17.84 | 1.21 | 1.24 | 1.85 |
| 10 | 250 | 2.08 | 1 | 0 | 34 | 0.62 | 2.16 | 1.41 | 2.17 | 1.81 | 1.19 | 1.35 | 0.83 | 1.13 | 1.28 |
| 10 | 250 | 2.08 | 1 | 1 | 34 | 0.62 | 1.21 | 1.42 | 2.67 | 1.21 | 1.33 | 1.37 | 0.85 | 1.18 | 1.36 |
| 10 | 500 | 2.08 | 1 | 0 | 11 | 0.47 | 2.11 | 1.41 | 1.49 | 1.75 | 1.35 | 1.04 | 0.65 | 0.95 | 0.94 |
| 10 | 500 | 2.08 | 1 | 1 | 11 | 0.49 | 1.47 | 1.4 | 1.47 | 1.38 | 1.21 | 1.04 | 0.68 | 0.99 | 0.97 |
| 2 | 100 | 1.04 | 0.5 | 0 | 47 | 0.5 | 6.6 | 1.13 | 0.93 | 6.65 | 3.33 | 1 | 0.71 | 1.14 | 0.91 |
| 2 | 100 | 1.04 | 0.5 | 1 | 47 | 0.5 | 0.86 | 0.9 | 2.8 | 0.86 | 0.9 | 3.61 | 0.74 | 1.2 | 1.01 |
| 2 | 250 | 1.04 | 0.5 | 0 | 16 | 0.41 | 4.93 | 1.73 | 0.86 | 4.96 | 3.52 | 1.35 | 0.62 | 0.94 | 0.75 |
| 2 | 250 | 1.04 | 0.5 | 1 | 16 | 0.42 | 1.22 | 0.8 | 0.9 | 0.85 | 0.78 | 0.68 | 0.64 | 0.97 | 0.78 |
| 2 | 500 | 1.04 | 0.5 | 0 | 2 | 0.31 | 2.14 | 0.97 | 0.74 | 2.11 | 1.51 | 0.86 | 0.52 | 0.68 | 0.6 |
| 2 | 500 | 1.04 | 0.5 | 1 | 2 | 0.32 | 1.49 | 0.7 | 0.52 | 1.47 | 1.01 | 0.63 | 0.53 | 0.7 | 0.61 |
| 5 | 100 | 1.04 | 0.5 | 0 | 50 | 0.5 | 0.9 | 0.85 | 1.02 | 0.84 | 0.85 | 0.73 | 0.7 | 1.13 | 0.93 |
| 5 | 100 | 1.04 | 0.5 | 1 | 50 | 0.49 | 0.85 | 0.9 | 2.51 | 0.84 | 0.9 | 1.92 | 0.73 | 1.22 | 1.03 |
| 5 | 250 | 1.04 | 0.5 | 0 | 18 | 0.39 | 3.17 | 0.71 | 0.71 | 2.88 | 1.11 | 0.63 | 0.62 | 0.98 | 0.76 |
| 5 | 250 | 1.04 | 0.5 | 1 | 18 | 0.39 | 0.73 | 0.81 | 1.11 | 0.72 | 0.78 | 0.67 | 0.64 | 1.02 | 0.8 |
| 5 | 500 | 1.04 | 0.5 | 0 | 3 | 0.33 | 1.67 | 0.94 | 0.58 | 1.69 | 1.58 | 0.82 | 0.54 | 0.72 | 0.63 |
| 5 | 500 | 1.04 | 0.5 | 1 | 3 | 0.34 | 1.38 | 0.71 | 0.61 | 1.15 | 0.66 | 0.59 | 0.56 | 0.74 | 0.64 |
| 10 | 100 | 1.04 | 0.5 | 0 | 48 | 0.49 | 0.82 | 0.88 | 4.8 | 0.81 | 0.87 | 2.4 | 0.75 | 1.29 | 1.25 |
| 10 | 100 | 1.04 | 0.5 | 1 | 48 | 0.48 | 0.84 | 0.9 | 11.1 | 0.84 | 0.89 | 10.82 | 0.76 | 1.38 | 1.36 |
| 10 | 250 | 1.04 | 0.5 | 0 | 12 | 0.41 | 1.02 | 0.75 | 1.21 | 1.01 | 0.72 | 0.65 | 0.67 | 1.03 | 0.87 |
| 10 | 250 | 1.04 | 0.5 | 1 | 12 | 0.41 | 0.71 | 0.8 | 1.33 | 0.71 | 0.76 | 0.67 | 0.68 | 1.06 | 0.89 |
| 10 | 500 | 1.04 | 0.5 | 0 | 2 | 0.32 | 1.19 | 0.62 | 0.67 | 1.02 | 0.57 | 0.58 | 0.56 | 0.74 | 0.66 |
| 10 | 500 | 1.04 | 0.5 | 1 | 2 | 0.33 | 0.58 | 0.68 | 0.72 | 0.57 | 0.62 | 0.58 | 0.57 | 0.76 | 0.67 |

Table S2. Simulation results showing prevalence of separation (SP, %) and root mean squared errors of $\beta_1$ across simulation scenarios with marginal event rate $E(Y) = 0.25$ that differed by the number of predictors $K \in \{2, 5, 10\}$, the sample size $N \in \{100, 250, 500\}$, the effect multiplier $a \in \{0.5, 1\}$ and noise absent (0) or present (1). OEX, explanation oracle; *D*, deviance; *GCV*, generalized cross-validation; *CE*, classification error; RCV50, repeated 10-fold cross-validated deviance with $\theta = 0.5$; RCV95, repeated 10-fold cross-validated deviance with $\theta = 0.95$; AIC, Akaike's information criterion; IP, shrinkage based on informative priors; WP, shrinkage based on weakly informative priors; FC, Firth's correction.

| *K* | *N* | $\beta_1$ | *a* | Noise | SP | OEX | *D* | *GCV* | *CE* | RCV50 | RCV95 | AIC | IP | WP | FC |
|---|---|---|---|---|---|---|---|---|---|---|---|---|---|---|---|
| 2 | 100 | 2.08 | 1 | 0 | 47 | 0.62 | 6.75 | 1.69 | 1.71 | 6.84 | 6.18 | 1.77 | 0.77 | 0.99 | 0.86 |
| 2 | 100 | 2.08 | 1 | 1 | 47 | 0.61 | 1.25 | 1.52 | 1.27 | 1.18 | 1.33 | 1.18 | 0.78 | 1.04 | 0.91 |
| 2 | 250 | 2.08 | 1 | 0 | 14 | 0.41 | 3.55 | 1.65 | 1.57 | 3.64 | 3.59 | 1.57 | 0.57 | 0.83 | 0.72 |
| 2 | 250 | 2.08 | 1 | 1 | 14 | 0.4 | 3.45 | 1.14 | 1.1 | 3.39 | 2.02 | 1.23 | 0.57 | 0.85 | 0.73 |
| 2 | 500 | 2.08 | 1 | 0 | 3 | 0.32 | 1.63 | 0.99 | 1.41 | 1.65 | 1.66 | 0.97 | 0.49 | 0.65 | 0.6 |
| 2 | 500 | 2.08 | 1 | 1 | 3 | 0.31 | 1.68 | 1.11 | 0.98 | 1.7 | 1.69 | 1.01 | 0.5 | 0.67 | 0.61 |
| 5 | 100 | 2.08 | 1 | 0 | 53 | 0.63 | 2.09 | 1.29 | 1.31 | 1.71 | 1.11 | 1.03 | 0.77 | 1 | 0.9 |
| 5 | 100 | 2.08 | 1 | 1 | 53 | 0.63 | 1.26 | 1.58 | 1.38 | 1.26 | 1.39 | 1.23 | 0.81 | 1.06 | 0.97 |
| 5 | 250 | 2.08 | 1 | 0 | 23 | 0.43 | 4.51 | 1.15 | 1.18 | 4.54 | 4.22 | 1.64 | 0.6 | 0.94 | 0.79 |
| 5 | 250 | 2.08 | 1 | 1 | 23 | 0.43 | 2.03 | 1.23 | 0.99 | 1.39 | 0.98 | 0.95 | 0.61 | 0.96 | 0.81 |
| 5 | 500 | 2.08 | 1 | 0 | 5 | 0.34 | 2.15 | 1.18 | 1.06 | 2.17 | 2.13 | 1.13 | 0.51 | 0.71 | 0.64 |
| 5 | 500 | 2.08 | 1 | 1 | 5 | 0.34 | 2.11 | 0.95 | 0.83 | 2.12 | 2.06 | 1.13 | 0.52 | 0.72 | 0.65 |
| 10 | 100 | 2.08 | 1 | 0 | 30 | 0.65 | 1.31 | 1.42 | 1.8 | 1.31 | 1.21 | 1.27 | 0.86 | 1.18 | 1.21 |
| 10 | 100 | 2.08 | 1 | 1 | 30 | 0.63 | 1.26 | 1.58 | 2.38 | 1.23 | 1.36 | 1.73 | 0.88 | 1.28 | 1.27 |
| 10 | 250 | 2.08 | 1 | 0 | 4 | 0.41 | 0.93 | 0.98 | 0.99 | 0.92 | 0.82 | 0.86 | 0.61 | 0.83 | 0.8 |
| 10 | 250 | 2.08 | 1 | 1 | 4 | 0.42 | 0.87 | 1.18 | 1.03 | 0.87 | 0.95 | 0.92 | 0.63 | 0.88 | 0.84 |
| 10 | 500 | 2.08 | 1 | 0 | 0 | 0.29 | 0.62 | 0.67 | 0.65 | 0.62 | 0.57 | 0.58 | 0.46 | 0.56 | 0.55 |
| 10 | 500 | 2.08 | 1 | 1 | 0 | 0.29 | 0.61 | 0.84 | 0.66 | 0.6 | 0.66 | 0.63 | 0.47 | 0.58 | 0.56 |
| 2 | 100 | 1.04 | 0.5 | 0 | 12 | 0.42 | 3.68 | 1.22 | 0.82 | 3.75 | 3.47 | 1.28 | 0.65 | 0.95 | 0.78 |
| 2 | 100 | 1.04 | 0.5 | 1 | 12 | 0.42 | 0.76 | 0.83 | 0.84 | 0.75 | 0.8 | 0.69 | 0.67 | 0.99 | 0.81 |
| 2 | 250 | 1.04 | 0.5 | 0 | 0 | 0.29 | 0.65 | 0.58 | 0.71 | 0.65 | 0.67 | 0.56 | 0.47 | 0.55 | 0.51 |
| 2 | 250 | 1.04 | 0.5 | 1 | 0 | 0.29 | 0.72 | 0.69 | 0.53 | 0.72 | 0.69 | 0.57 | 0.49 | 0.58 | 0.53 |
| 2 | 500 | 1.04 | 0.5 | 0 | 0 | 0.23 | 0.36 | 0.4 | 0.59 | 0.36 | 0.37 | 0.37 | 0.35 | 0.37 | 0.36 |
| 2 | 500 | 1.04 | 0.5 | 1 | 0 | 0.23 | 0.43 | 0.56 | 0.46 | 0.43 | 0.47 | 0.45 | 0.35 | 0.38 | 0.37 |
| 5 | 100 | 1.04 | 0.5 | 0 | 14 | 0.39 | 1.9 | 0.75 | 0.75 | 1.5 | 0.71 | 0.65 | 0.66 | 1 | 0.82 |
| 5 | 100 | 1.04 | 0.5 | 1 | 14 | 0.4 | 0.73 | 0.83 | 0.93 | 0.73 | 0.78 | 0.66 | 0.69 | 1.08 | 0.87 |
| 5 | 250 | 1.04 | 0.5 | 0 | 1 | 0.3 | 0.97 | 0.63 | 0.56 | 0.98 | 0.98 | 0.61 | 0.52 | 0.63 | 0.57 |
| 5 | 250 | 1.04 | 0.5 | 1 | 1 | 0.3 | 0.77 | 0.7 | 0.56 | 0.65 | 0.61 | 0.55 | 0.54 | 0.65 | 0.59 |
| 5 | 500 | 1.04 | 0.5 | 0 | 0 | 0.22 | 0.36 | 0.44 | 0.46 | 0.36 | 0.37 | 0.37 | 0.37 | 0.4 | 0.39 |
| 5 | 500 | 1.04 | 0.5 | 1 | 0 | 0.22 | 0.42 | 0.57 | 0.39 | 0.42 | 0.46 | 0.44 | 0.38 | 0.41 | 0.39 |
| 10 | 100 | 1.04 | 0.5 | 0 | 10 | 0.43 | 0.71 | 0.82 | 0.96 | 0.71 | 0.76 | 0.67 | 0.72 | 1.08 | 0.92 |
| 10 | 100 | 1.04 | 0.5 | 1 | 10 | 0.44 | 0.75 | 0.86 | 1.37 | 0.75 | 0.8 | 0.69 | 0.76 | 1.2 | 1.03 |
| 10 | 250 | 1.04 | 0.5 | 0 | 0 | 0.32 | 0.54 | 0.67 | 0.58 | 0.53 | 0.57 | 0.54 | 0.51 | 0.6 | 0.56 |
| 10 | 250 | 1.04 | 0.5 | 1 | 0 | 0.32 | 0.58 | 0.73 | 0.59 | 0.58 | 0.62 | 0.57 | 0.53 | 0.63 | 0.58 |
| 10 | 500 | 1.04 | 0.5 | 0 | 0 | 0.25 | 0.4 | 0.53 | 0.41 | 0.39 | 0.42 | 0.42 | 0.37 | 0.4 | 0.39 |
| 10 | 500 | 1.04 | 0.5 | 1 | 0 | 0.26 | 0.44 | 0.6 | 0.43 | 0.44 | 0.48 | 0.47 | 0.39 | 0.42 | 0.41 |

Table S3. Simulation results showing prevalence of separation (SP, %) and root mean squared errors of $\beta_2$ across simulation scenarios with marginal event rate $E(Y) = 0.10$ that differed by the number of predictors $K \in \{2, 5, 10\}$, the sample size $N \in \{100, 250, 500\}$, the effect multiplier $a \in \{0.5, 1\}$ and noise absent (0) or present (1). OEX, explanation oracle; *D*, deviance; *GCV*, generalized cross-validation; *CE*, classification error; RCV50, repeated 10-fold cross-validated deviance with $\theta = 0.5$; RCV95, repeated 10-fold cross-validated deviance with $\theta = 0.95$; AIC, Akaike's information criterion; IP, shrinkage based on informative priors; WP, shrinkage based on weakly informative priors; FC, Firth's correction.

| $K$ | $N$ | $\beta_2$ | $a$ | Noise | SP | OEX | D | GCV | CE | RCV50 | RCV95 | AIC | IP | WP | FC |
|---|---|---|---|---|---|---|---|---|---|---|---|---|---|---|---|
| 2 | 100 | 1.39 | 1 | 0 | 79 | 1.29 | 2.95 | 1.3 | 1.36 | 2.95 | 2.73 | 1 | 0.62 | 0.8 | 0.77 |
| 2 | 100 | 1.39 | 1 | 1 | 79 | 1.51 | 0.91 | 0.99 | 2.83 | 0.9 | 0.96 | 4.72 | 0.65 | 0.91 | 0.9 |
| 2 | 250 | 1.39 | 1 | 0 | 54 | 0.48 | 0.81 | 0.57 | 0.93 | 0.81 | 0.8 | 0.53 | 0.43 | 0.49 | 0.48 |
| 2 | 250 | 1.39 | 1 | 1 | 54 | 0.51 | 0.63 | 0.69 | 0.55 | 0.56 | 0.63 | 0.56 | 0.44 | 0.52 | 0.49 |
| 2 | 500 | 1.39 | 1 | 0 | 33 | 0.33 | 0.33 | 0.34 | 0.81 | 0.33 | 0.33 | 0.33 | 0.32 | 0.34 | 0.33 |
| 2 | 500 | 1.39 | 1 | 1 | 33 | 0.34 | 0.37 | 0.45 | 0.42 | 0.36 | 0.39 | 0.38 | 0.32 | 0.34 | 0.34 |
| 5 | 100 | 1.39 | 1 | 0 | 85 | 1.43 | 1.25 | 0.9 | 0.95 | 1.02 | 0.88 | 0.84 | 0.63 | 0.82 | 0.77 |
| 5 | 100 | 1.39 | 1 | 1 | 85 | 1.78 | 0.93 | 1.04 | 3.87 | 0.92 | 0.99 | 5.24 | 0.65 | 0.93 | 0.87 |
| 5 | 250 | 1.39 | 1 | 0 | 63 | 0.5 | 0.49 | 0.56 | 0.6 | 0.49 | 0.52 | 0.5 | 0.45 | 0.51 | 0.49 |
| 5 | 250 | 1.39 | 1 | 1 | 63 | 0.53 | 0.6 | 0.75 | 0.56 | 0.6 | 0.68 | 0.59 | 0.47 | 0.55 | 0.51 |
| 5 | 500 | 1.39 | 1 | 0 | 41 | 0.33 | 0.33 | 0.35 | 0.45 | 0.33 | 0.33 | 0.33 | 0.31 | 0.34 | 0.33 |
| 5 | 500 | 1.39 | 1 | 1 | 41 | 0.35 | 0.37 | 0.5 | 0.36 | 0.37 | 0.42 | 0.41 | 0.33 | 0.35 | 0.34 |
| 10 | 100 | 1.39 | 1 | 0 | 72 | 2.5 | 0.95 | 1.01 | 8.74 | 0.92 | 0.92 | 7.59 | 0.69 | 0.99 | 1.04 |
| 10 | 100 | 1.39 | 1 | 1 | 72 | 3.37 | 0.96 | 1.04 | 13.35 | 0.92 | 0.97 | 10.66 | 0.71 | 1.09 | 0.99 |
| 10 | 250 | 1.39 | 1 | 0 | 34 | 0.58 | 0.53 | 0.6 | 0.65 | 0.53 | 0.56 | 0.55 | 0.47 | 0.59 | 0.55 |
| 10 | 250 | 1.39 | 1 | 1 | 34 | 0.64 | 0.58 | 0.69 | 0.75 | 0.57 | 0.63 | 0.59 | 0.49 | 0.64 | 0.58 |
| 10 | 500 | 1.39 | 1 | 0 | 11 | 0.39 | 0.36 | 0.4 | 0.41 | 0.36 | 0.37 | 0.37 | 0.35 | 0.39 | 0.38 |
| 10 | 500 | 1.39 | 1 | 1 | 11 | 0.4 | 0.39 | 0.47 | 0.42 | 0.39 | 0.42 | 0.4 | 0.36 | 0.41 | 0.38 |
| 2 | 100 | 0.69 | 0.5 | 0 | 47 | 0.62 | 1.66 | 0.86 | 0.62 | 1.67 | 1.59 | 0.71 | 0.6 | 0.75 | 0.71 |
| 2 | 100 | 0.69 | 0.5 | 1 | 47 | 2.64 | 0.62 | 0.6 | 2.44 | 0.62 | 0.6 | 4.89 | 0.63 | 0.85 | 0.8 |
| 2 | 250 | 0.69 | 0.5 | 0 | 16 | 0.43 | 0.44 | 0.44 | 0.52 | 0.44 | 0.45 | 0.43 | 0.43 | 0.47 | 0.46 |
| 2 | 250 | 0.69 | 0.5 | 1 | 16 | 0.45 | 0.45 | 0.48 | 0.46 | 0.45 | 0.48 | 0.42 | 0.45 | 0.5 | 0.48 |
| 2 | 500 | 0.69 | 0.5 | 0 | 2 | 0.3 | 0.3 | 0.31 | 0.43 | 0.3 | 0.31 | 0.3 | 0.31 | 0.32 | 0.31 |
| 2 | 500 | 0.69 | 0.5 | 1 | 2 | 0.31 | 0.34 | 0.38 | 0.32 | 0.34 | 0.36 | 0.33 | 0.31 | 0.33 | 0.32 |
| 5 | 100 | 0.69 | 0.5 | 0 | 50 | 0.61 | 0.58 | 0.57 | 0.73 | 0.58 | 0.58 | 0.56 | 0.6 | 0.76 | 0.71 |
| 5 | 100 | 0.69 | 0.5 | 1 | 50 | 0.73 | 0.59 | 0.59 | 3.47 | 0.59 | 0.59 | 5.56 | 0.63 | 0.88 | 0.79 |
| 5 | 250 | 0.69 | 0.5 | 0 | 18 | 0.4 | 0.43 | 0.45 | 0.41 | 0.42 | 0.44 | 0.39 | 0.41 | 0.44 | 0.43 |
| 5 | 250 | 0.69 | 0.5 | 1 | 18 | 0.42 | 0.45 | 0.5 | 0.44 | 0.45 | 0.48 | 0.41 | 0.42 | 0.47 | 0.44 |
| 5 | 500 | 0.69 | 0.5 | 0 | 3 | 0.31 | 0.32 | 0.35 | 0.34 | 0.32 | 0.34 | 0.31 | 0.32 | 0.33 | 0.32 |
| 5 | 500 | 0.69 | 0.5 | 1 | 3 | 0.33 | 0.36 | 0.41 | 0.33 | 0.36 | 0.39 | 0.35 | 0.33 | 0.34 | 0.33 |
| 10 | 100 | 0.69 | 0.5 | 0 | 48 | 3.56 | 0.65 | 0.64 | 5.08 | 0.62 | 0.58 | 4 | 0.66 | 0.94 | 0.85 |
| 10 | 100 | 0.69 | 0.5 | 1 | 48 | 3.23 | 0.59 | 0.59 | 9.37 | 0.59 | 0.58 | 8.93 | 0.68 | 1.06 | 0.94 |
| 10 | 250 | 0.69 | 0.5 | 0 | 12 | 0.42 | 0.41 | 0.45 | 0.46 | 0.41 | 0.43 | 0.39 | 0.42 | 0.47 | 0.45 |
| 10 | 250 | 0.69 | 0.5 | 1 | 12 | 0.45 | 0.43 | 0.48 | 0.52 | 0.43 | 0.46 | 0.4 | 0.44 | 0.52 | 0.47 |
| 10 | 500 | 0.69 | 0.5 | 0 | 2 | 0.31 | 0.3 | 0.34 | 0.34 | 0.3 | 0.32 | 0.3 | 0.32 | 0.34 | 0.32 |
| 10 | 500 | 0.69 | 0.5 | 1 | 2 | 0.33 | 0.32 | 0.38 | 0.35 | 0.32 | 0.34 | 0.32 | 0.33 | 0.36 | 0.34 |

Table S4. Simulation results showing prevalence of separation (SP, %) and root mean squared errors of $\beta_2$ across simulation scenarios with marginal event rate $E(Y) = 0.25$ that differed by the number of predictors $K \in \{2, 5, 10\}$, the sample size $N \in \{100, 250, 500\}$, the effect multiplier $a \in \{0.5, 1\}$ and noise absent (0) or present (1). OEX, explanation oracle; *D*, deviance; *GCV*, generalized cross-validation; *CE*, classification error; RCV50, repeated 10-fold cross-validated deviance with $\theta = 0.5$; RCV95, repeated 10-fold cross-validated deviance with $\theta = 0.95$; AIC, Akaike's information criterion; IP, shrinkage based on informative priors; WP, shrinkage based on weakly informative priors; FC, Firth's correction.

| *K* | *N* | $\beta_2$ | *a* | Noise | SP | OEX | *D* | *GCV* | *CE* | RCV50 | RCV95 | AIC | IP | WP | FC |
|---|---|---|---|---|---|---|---|---|---|---|---|---|---|---|---|
| 2 | 100 | 1.39 | 1 | 0 | 47 | 0.51 | 0.52 | 0.54 | 1.02 | 0.52 | 0.52 | 0.52 | 0.47 | 0.52 | 0.51 |
| 2 | 100 | 1.39 | 1 | 1 | 47 | 0.57 | 0.6 | 0.79 | 0.63 | 0.59 | 0.67 | 0.59 | 0.49 | 0.58 | 0.54 |
| 2 | 250 | 1.39 | 1 | 0 | 14 | 0.33 | 0.32 | 0.34 | 0.83 | 0.32 | 0.33 | 0.33 | 0.32 | 0.33 | 0.33 |
| 2 | 250 | 1.39 | 1 | 1 | 14 | 0.34 | 0.37 | 0.5 | 0.52 | 0.37 | 0.4 | 0.39 | 0.33 | 0.35 | 0.34 |
| 2 | 500 | 1.39 | 1 | 0 | 3 | 0.22 | 0.22 | 0.23 | 0.66 | 0.22 | 0.22 | 0.22 | 0.22 | 0.22 | 0.22 |
| 2 | 500 | 1.39 | 1 | 1 | 3 | 0.23 | 0.24 | 0.31 | 0.41 | 0.24 | 0.25 | 0.25 | 0.23 | 0.23 | 0.23 |
| 5 | 100 | 1.39 | 1 | 0 | 53 | 0.52 | 0.52 | 0.66 | 0.69 | 0.52 | 0.57 | 0.54 | 0.46 | 0.53 | 0.51 |
| 5 | 100 | 1.39 | 1 | 1 | 53 | 0.59 | 0.65 | 0.87 | 0.64 | 0.65 | 0.73 | 0.63 | 0.49 | 0.62 | 0.57 |
| 5 | 250 | 1.39 | 1 | 0 | 23 | 0.33 | 0.33 | 0.38 | 0.58 | 0.33 | 0.34 | 0.33 | 0.31 | 0.33 | 0.33 |
| 5 | 250 | 1.39 | 1 | 1 | 23 | 0.36 | 0.38 | 0.57 | 0.43 | 0.38 | 0.43 | 0.41 | 0.33 | 0.36 | 0.34 |
| 5 | 500 | 1.39 | 1 | 0 | 5 | 0.24 | 0.24 | 0.26 | 0.47 | 0.24 | 0.24 | 0.24 | 0.24 | 0.24 | 0.24 |
| 5 | 500 | 1.39 | 1 | 1 | 5 | 0.25 | 0.26 | 0.37 | 0.34 | 0.26 | 0.28 | 0.28 | 0.24 | 0.25 | 0.25 |
| 10 | 100 | 1.39 | 1 | 0 | 30 | 0.69 | 0.59 | 0.75 | 0.72 | 0.58 | 0.63 | 0.64 | 0.52 | 0.68 | 0.63 |
| 10 | 100 | 1.39 | 1 | 1 | 30 | 0.81 | 0.65 | 0.87 | 1.14 | 0.65 | 0.71 | 1.09 | 0.55 | 0.8 | 0.72 |
| 10 | 250 | 1.39 | 1 | 0 | 4 | 0.38 | 0.36 | 0.45 | 0.43 | 0.36 | 0.38 | 0.37 | 0.35 | 0.39 | 0.37 |
| 10 | 250 | 1.39 | 1 | 1 | 4 | 0.41 | 0.4 | 0.55 | 0.44 | 0.4 | 0.43 | 0.41 | 0.36 | 0.42 | 0.39 |
| 10 | 500 | 1.39 | 1 | 0 | 0 | 0.26 | 0.25 | 0.28 | 0.3 | 0.25 | 0.25 | 0.25 | 0.25 | 0.26 | 0.25 |
| 10 | 500 | 1.39 | 1 | 1 | 0 | 0.27 | 0.27 | 0.35 | 0.3 | 0.27 | 0.28 | 0.28 | 0.26 | 0.27 | 0.26 |
| 2 | 100 | 0.69 | 0.5 | 0 | 12 | 0.46 | 0.46 | 0.48 | 0.53 | 0.47 | 0.48 | 0.45 | 0.46 | 0.51 | 0.5 |
| 2 | 100 | 0.69 | 0.5 | 1 | 12 | 0.48 | 0.49 | 0.53 | 0.47 | 0.49 | 0.51 | 0.45 | 0.5 | 0.56 | 0.53 |
| 2 | 250 | 0.69 | 0.5 | 0 | 0 | 0.29 | 0.3 | 0.33 | 0.43 | 0.3 | 0.31 | 0.3 | 0.3 | 0.31 | 0.31 |
| 2 | 250 | 0.69 | 0.5 | 1 | 0 | 0.31 | 0.35 | 0.41 | 0.32 | 0.35 | 0.37 | 0.34 | 0.32 | 0.33 | 0.32 |
| 2 | 500 | 0.69 | 0.5 | 0 | 0 | 0.22 | 0.21 | 0.23 | 0.32 | 0.21 | 0.22 | 0.21 | 0.22 | 0.22 | 0.22 |
| 2 | 500 | 0.69 | 0.5 | 1 | 0 | 0.22 | 0.24 | 0.3 | 0.26 | 0.24 | 0.26 | 0.25 | 0.23 | 0.23 | 0.23 |
| 5 | 100 | 0.69 | 0.5 | 0 | 14 | 0.46 | 0.46 | 0.49 | 0.46 | 0.46 | 0.47 | 0.43 | 0.47 | 0.53 | 0.51 |
| 5 | 100 | 0.69 | 0.5 | 1 | 14 | 0.51 | 0.48 | 0.53 | 0.52 | 0.48 | 0.51 | 0.44 | 0.51 | 0.6 | 0.55 |
| 5 | 250 | 0.69 | 0.5 | 0 | 1 | 0.29 | 0.3 | 0.36 | 0.34 | 0.3 | 0.32 | 0.3 | 0.3 | 0.31 | 0.31 |
| 5 | 250 | 0.69 | 0.5 | 1 | 1 | 0.31 | 0.34 | 0.42 | 0.31 | 0.34 | 0.37 | 0.33 | 0.31 | 0.33 | 0.31 |
| 5 | 500 | 0.69 | 0.5 | 0 | 0 | 0.22 | 0.21 | 0.25 | 0.27 | 0.21 | 0.22 | 0.22 | 0.22 | 0.22 | 0.22 |
| 5 | 500 | 0.69 | 0.5 | 1 | 0 | 0.23 | 0.25 | 0.33 | 0.24 | 0.25 | 0.27 | 0.26 | 0.23 | 0.23 | 0.23 |
| 10 | 100 | 0.69 | 0.5 | 0 | 10 | 0.51 | 0.46 | 0.51 | 0.52 | 0.46 | 0.48 | 0.44 | 0.51 | 0.61 | 0.55 |
| 10 | 100 | 0.69 | 0.5 | 1 | 10 | 0.56 | 0.47 | 0.54 | 0.57 | 0.47 | 0.5 | 0.44 | 0.55 | 0.7 | 0.61 |
| 10 | 250 | 0.69 | 0.5 | 0 | 0 | 0.31 | 0.31 | 0.38 | 0.32 | 0.31 | 0.33 | 0.31 | 0.32 | 0.34 | 0.32 |
| 10 | 250 | 0.69 | 0.5 | 1 | 0 | 0.33 | 0.33 | 0.42 | 0.33 | 0.33 | 0.35 | 0.32 | 0.33 | 0.36 | 0.33 |
| 10 | 500 | 0.69 | 0.5 | 0 | 0 | 0.23 | 0.22 | 0.28 | 0.24 | 0.22 | 0.23 | 0.23 | 0.23 | 0.23 | 0.23 |
| 10 | 500 | 0.69 | 0.5 | 1 | 0 | 0.24 | 0.24 | 0.32 | 0.24 | 0.24 | 0.25 | 0.25 | 0.24 | 0.24 | 0.24 |

Table S5. Simulation results showing root mean squared errors of predictions ($\times 10000$) across simulation scenarios with marginal event rate $\mathrm{E}(Y) = 0.10$ that differed by the number of predictors $K \in \{2, 5, 10\}$, the sample size $N \in \{100, 250, 500\}$, the effect multiplier $a \in \{0.5, 1\}$ and noise absent (0) or present (1). OP, prediction oracle; *D*, deviance; *GCV*, generalized cross-validation; *CE*, classification error; RCV50, repeated 10-fold cross-validated deviance with $\theta = 0.5$; RCV95, repeated 10-fold cross-validated deviance with $\theta = 0.95$; AIC, Akaike's information criterion; IP, shrinkage based on informative priors; WP, shrinkage based on weakly informative priors; FLIC, Firth's logistic regression with intercept-correction.

| K | N | a | Noise | OP | D | GCV | CE | RCV50 | RCV95 | AIC | IP | WP | FLIC |
|---|---|---|---|---|---|---|---|---|---|---|---|---|---|
| 2 | 100 | 1 | 0 | 405 | 493 | 514 | 641 | 493 | 525 | 498 | 459 | 469 | 480 |
| 2 | 100 | 1 | 1 | 591 | 700 | 684 | 877 | 701 | 691 | 697 | 744 | 881 | 843 |
| 2 | 250 | 1 | 0 | 261 | 319 | 330 | 560 | 317 | 327 | 319 | 295 | 299 | 303 |
| 2 | 250 | 1 | 1 | 439 | 510 | 504 | 547 | 506 | 501 | 491 | 533 | 579 | 568 |
| 2 | 500 | 1 | 0 | 182 | 223 | 230 | 483 | 222 | 228 | 223 | 211 | 214 | 215 |
| 2 | 500 | 1 | 1 | 340 | 398 | 384 | 404 | 398 | 391 | 388 | 402 | 420 | 415 |
| 5 | 100 | 1 | 0 | 580 | 697 | 698 | 721 | 699 | 701 | 682 | 660 | 749 | 736 |
| 5 | 100 | 1 | 1 | 678 | 793 | 769 | 1131 | 793 | 778 | 827 | 881 | 1066 | 990 |
| 5 | 250 | 1 | 0 | 409 | 486 | 484 | 503 | 483 | 477 | 471 | 456 | 487 | 484 |
| 5 | 250 | 1 | 1 | 527 | 585 | 589 | 686 | 584 | 584 | 574 | 643 | 707 | 681 |
| 5 | 500 | 1 | 0 | 304 | 349 | 356 | 387 | 349 | 352 | 351 | 338 | 351 | 349 |
| 5 | 500 | 1 | 1 | 413 | 466 | 456 | 487 | 460 | 451 | 449 | 480 | 504 | 494 |
| 10 | 100 | 1 | 0 | 833 | 940 | 966 | 1216 | 938 | 958 | 1039 | 909 | 1068 | 1039 |
| 10 | 100 | 1 | 1 | 902 | 1011 | 1036 | 1591 | 1006 | 1018 | 1377 | 1037 | 1267 | 1185 |
| 10 | 250 | 1 | 0 | 617 | 672 | 692 | 739 | 669 | 675 | 678 | 664 | 725 | 716 |
| 10 | 250 | 1 | 1 | 693 | 741 | 767 | 893 | 739 | 746 | 749 | 779 | 873 | 846 |
| 10 | 500 | 1 | 0 | 471 | 506 | 518 | 530 | 505 | 507 | 510 | 499 | 525 | 521 |
| 10 | 500 | 1 | 1 | 545 | 575 | 592 | 635 | 575 | 577 | 578 | 594 | 633 | 621 |
| 2 | 100 | 0.5 | 0 | 381 | 483 | 468 | 467 | 483 | 480 | 459 | 440 | 472 | 470 |
| 2 | 100 | 0.5 | 1 | 463 | 559 | 514 | 848 | 559 | 526 | 570 | 720 | 857 | 810 |
| 2 | 250 | 0.5 | 0 | 258 | 319 | 331 | 384 | 318 | 334 | 315 | 293 | 304 | 304 |
| 2 | 250 | 0.5 | 1 | 357 | 417 | 398 | 515 | 415 | 404 | 398 | 510 | 553 | 540 |
| 2 | 500 | 0.5 | 0 | 194 | 232 | 244 | 321 | 232 | 242 | 235 | 220 | 225 | 225 |
| 2 | 500 | 0.5 | 1 | 291 | 331 | 326 | 367 | 332 | 329 | 319 | 382 | 398 | 394 |
| 5 | 100 | 0.5 | 0 | 472 | 566 | 538 | 632 | 567 | 550 | 553 | 634 | 733 | 708 |
| 5 | 100 | 0.5 | 1 | 508 | 608 | 559 | 1061 | 608 | 575 | 648 | 853 | 1033 | 939 |
| 5 | 250 | 0.5 | 0 | 343 | 418 | 402 | 407 | 417 | 408 | 389 | 434 | 465 | 458 |
| 5 | 250 | 0.5 | 1 | 398 | 456 | 435 | 636 | 456 | 444 | 443 | 609 | 667 | 639 |
| 5 | 500 | 0.5 | 0 | 275 | 314 | 323 | 324 | 315 | 321 | 310 | 328 | 340 | 337 |
| 5 | 500 | 0.5 | 1 | 332 | 373 | 364 | 450 | 372 | 366 | 358 | 453 | 474 | 465 |
| 10 | 100 | 0.5 | 0 | 664 | 773 | 762 | 1102 | 771 | 768 | 789 | 898 | 1077 | 980 |
| 10 | 100 | 0.5 | 1 | 701 | 818 | 792 | 1470 | 814 | 801 | 1045 | 1047 | 1294 | 1132 |
| 10 | 250 | 0.5 | 0 | 519 | 587 | 602 | 679 | 586 | 596 | 566 | 654 | 710 | 676 |
| 10 | 250 | 0.5 | 1 | 564 | 625 | 639 | 827 | 624 | 631 | 605 | 773 | 855 | 798 |
| 10 | 500 | 0.5 | 0 | 406 | 446 | 467 | 489 | 445 | 452 | 443 | 483 | 504 | 490 |
| 10 | 500 | 0.5 | 1 | 453 | 489 | 512 | 583 | 488 | 494 | 482 | 577 | 608 | 586 |

Table S6. Simulation results showing root mean squared errors of predictions ($\times 10000$) across simulation scenarios with marginal event rate $\mathrm{E}(Y) = 0.25$ that differed by the number of predictors $K \in \{2, 5, 10\}$, the sample size $N \in \{100, 250, 500\}$, the effect multiplier $a \in \{0.5, 1\}$ and noise absent (0) or present (1). OP, prediction oracle; *D*, deviance; *GCV*, generalized cross-validation; *CE*, classification error; RCV50, repeated 10-fold cross-validated deviance with $\theta = 0.5$; RCV95, repeated 10-fold cross-validated deviance with $\theta = 0.95$; AIC, Akaike's information criterion; IP, shrinkage based on informative priors; WP, shrinkage based on weakly informative priors; FLIC, Firth's logistic regression with intercept-correction.

| K | N | a | Noise | OP | D | GCV | CE | RCV50 | RCV95 | AIC | IP | WP | FLIC |
|---|---|---|---|---|---|---|---|---|---|---|---|---|---|
| 2 | 100 | 1 | 0 | 586 | 700 | 749 | 1289 | 696 | 720 | 698 | 645 | 654 | 663 |
| 2 | 100 | 1 | 1 | 921 | 1031 | 1109 | 1086 | 1029 | 1050 | 1016 | 1053 | 1153 | 1124 |
| 2 | 250 | 1 | 0 | 375 | 438 | 467 | 1045 | 435 | 451 | 439 | 416 | 419 | 422 |
| 2 | 250 | 1 | 1 | 642 | 710 | 758 | 787 | 709 | 711 | 711 | 711 | 740 | 730 |
| 2 | 500 | 1 | 0 | 270 | 307 | 323 | 829 | 306 | 316 | 309 | 298 | 299 | 300 |
| 2 | 500 | 1 | 1 | 485 | 524 | 554 | 622 | 524 | 527 | 529 | 528 | 540 | 536 |
| 5 | 100 | 1 | 0 | 834 | 946 | 1026 | 1076 | 943 | 960 | 944 | 900 | 966 | 961 |
| 5 | 100 | 1 | 1 | 1040 | 1141 | 1229 | 1229 | 1139 | 1157 | 1133 | 1211 | 1350 | 1298 |
| 5 | 250 | 1 | 0 | 562 | 621 | 659 | 841 | 620 | 629 | 627 | 603 | 624 | 623 |
| 5 | 250 | 1 | 1 | 750 | 814 | 872 | 854 | 811 | 810 | 810 | 837 | 879 | 861 |
| 5 | 500 | 1 | 0 | 417 | 451 | 475 | 688 | 451 | 456 | 456 | 445 | 453 | 452 |
| 5 | 500 | 1 | 1 | 576 | 611 | 648 | 649 | 611 | 613 | 615 | 620 | 635 | 628 |
| 10 | 100 | 1 | 0 | 1109 | 1187 | 1305 | 1292 | 1186 | 1207 | 1209 | 1178 | 1303 | 1287 |
| 10 | 100 | 1 | 1 | 1246 | 1324 | 1455 | 1524 | 1322 | 1339 | 1371 | 1390 | 1583 | 1530 |
| 10 | 250 | 1 | 0 | 783 | 825 | 891 | 882 | 824 | 834 | 834 | 817 | 857 | 851 |
| 10 | 250 | 1 | 1 | 915 | 954 | 1039 | 1021 | 954 | 961 | 963 | 979 | 1042 | 1021 |
| 10 | 500 | 1 | 0 | 569 | 595 | 627 | 640 | 595 | 600 | 599 | 591 | 607 | 604 |
| 10 | 500 | 1 | 1 | 681 | 705 | 752 | 738 | 705 | 709 | 709 | 714 | 737 | 729 |
| 2 | 100 | 0.5 | 0 | 591 | 734 | 773 | 842 | 732 | 765 | 719 | 670 | 699 | 698 |
| 2 | 100 | 0.5 | 1 | 784 | 896 | 881 | 1006 | 897 | 888 | 870 | 1070 | 1170 | 1120 |
| 2 | 250 | 0.5 | 0 | 386 | 463 | 513 | 659 | 461 | 486 | 470 | 435 | 444 | 443 |
| 2 | 250 | 0.5 | 1 | 584 | 656 | 676 | 682 | 656 | 660 | 640 | 737 | 766 | 750 |
| 2 | 500 | 0.5 | 0 | 277 | 324 | 357 | 505 | 323 | 335 | 332 | 314 | 318 | 318 |
| 2 | 500 | 0.5 | 1 | 446 | 490 | 521 | 506 | 489 | 495 | 487 | 530 | 540 | 534 |
| 5 | 100 | 0.5 | 0 | 748 | 885 | 883 | 886 | 881 | 878 | 844 | 931 | 1011 | 988 |
| 5 | 100 | 0.5 | 1 | 852 | 962 | 935 | 1197 | 963 | 945 | 951 | 1254 | 1395 | 1312 |
| 5 | 250 | 0.5 | 0 | 541 | 617 | 659 | 652 | 617 | 630 | 612 | 641 | 664 | 657 |
| 5 | 250 | 0.5 | 1 | 662 | 729 | 742 | 810 | 728 | 727 | 711 | 869 | 908 | 881 |
| 5 | 500 | 0.5 | 0 | 396 | 438 | 478 | 500 | 438 | 444 | 442 | 453 | 462 | 459 |
| 5 | 500 | 0.5 | 1 | 511 | 551 | 581 | 579 | 551 | 554 | 546 | 618 | 632 | 622 |
| 10 | 100 | 0.5 | 0 | 1003 | 1108 | 1151 | 1241 | 1108 | 1119 | 1082 | 1265 | 1399 | 1324 |
| 10 | 100 | 0.5 | 1 | 1078 | 1178 | 1204 | 1454 | 1178 | 1180 | 1170 | 1490 | 1685 | 1554 |
| 10 | 250 | 0.5 | 0 | 722 | 786 | 870 | 829 | 785 | 801 | 781 | 842 | 879 | 858 |
| 10 | 250 | 0.5 | 1 | 813 | 869 | 943 | 962 | 868 | 880 | 857 | 1012 | 1065 | 1023 |
| 10 | 500 | 0.5 | 0 | 565 | 600 | 670 | 627 | 600 | 608 | 606 | 625 | 638 | 631 |
| 10 | 500 | 0.5 | 1 | 649 | 681 | 753 | 723 | 681 | 688 | 682 | 749 | 768 | 753 |

Table S7. Simulation results showing median calibration slopes (with 5th and 95th percentile) across simulation scenarios with marginal event rate $E(Y) = 0.10$ that differed by the number of predictors $K \in \{2, 5, 10\}$, the sample size $N \in \{100, 250, 500\}$, the effect multiplier $a \in \{0.5, 1\}$ and noise absent (0) or present (1). OP, prediction oracle; *D*, deviance; *GCV*, generalized cross-validation; *CE*, classification error; RCV50, repeated 10-fold cross-validated deviance with $\theta = 0.5$; RCV95, repeated 10-fold cross-validated deviance with $\theta = 0.95$; AIC, Akaike's information criterion; IP, shrinkage based on informative priors; WP, shrinkage based on weakly informative priors; FLIC, Firth's logistic regression with intercept-correction.

| K | N | a | Noise | OP | D | GCV | CE | RCV50 | RCV95 | AIC | IP | WP | FLIC |
|---|---|---|---|---|---|---|---|---|---|---|---|---|---|
| 2 | 100 | 1 | 0 | 1.1 (0.3, 2.1) | 0.7 (0.2, 5.7) | 1.4 (0.5, 15.3) | 4.9 (0.8, 11.5) | 0.5 (0.2, 4.8) | 1.3 (0.2, 28.1) | 1.2 (0.4, 5.1) | 1.2 (0.7, 2.8) | 1 (0.5, 2.1) | 1.1 (0.5, 2.6) |
| 2 | 100 | 1 | 1 | 1.2 (0.7, 1.7) | 1.2 (0.4, 11.5) | 1.8 (0.5, 12.1) | 0.5 (0.2, 1.7) | 1.1 (0.4, 11.3) | 1.6 (0.5, 12.7) | 1 (0.3, 2.9) | 0.7 (0.4, 1.1) | 0.5 (0.2, 0.8) | 0.5 (0.2, 1) |
| 2 | 250 | 1 | 0 | 1.1 (0.3, 1.6) | 0.4 (0.2, 2.1) | 0.8 (0.6, 2.5) | 4.3 (1, 8.7) | 0.4 (0.2, 2.1) | 0.8 (0.3, 2.4) | 0.9 (0.6, 2.2) | 1.1 (0.7, 1.8) | 1 (0.6, 1.6) | 1 (0.7, 1.7) |
| 2 | 250 | 1 | 1 | 1.1 (0.9, 1.4) | 1 (0.5, 2.3) | 1.4 (0.8, 3.6) | 0.8 (0.5, 2) | 1 (0.6, 2.3) | 1.2 (0.7, 3.3) | 1.1 (0.7, 2.2) | 0.8 (0.6, 1.1) | 0.7 (0.5, 1) | 0.7 (0.5, 1.1) |
| 2 | 500 | 1 | 0 | 1 (0.3, 1.3) | 1 (0.2, 1.5) | 1 (0.5, 1.6) | 2.8 (1.1, 4.3) | 0.9 (0.2, 1.5) | 1 (0.4, 1.5) | 0.9 (0.6, 1.5) | 1 (0.7, 1.4) | 0.9 (0.7, 1.3) | 0.9 (0.7, 1.3) |
| 2 | 500 | 1 | 1 | 1.1 (0.9, 1.2) | 0.9 (0.3, 1.4) | 1.1 (0.7, 1.8) | 1 (0.6, 1.9) | 0.9 (0.3, 1.4) | 1 (0.3, 1.6) | 1 (0.6, 1.5) | 0.8 (0.6, 1.1) | 0.8 (0.6, 1) | 0.8 (0.6, 1.1) |
| 5 | 100 | 1 | 0 | 1.1 (0.6, 1.6) | 1.1 (0.4, 15.4) | 1.6 (0.5, 17.1) | 0.9 (0.3, 4) | 1.1 (0.4, 14.9) | 1.4 (0.5, 17.4) | 1.1 (0.4, 3.5) | 0.9 (0.5, 1.5) | 0.7 (0.3, 1.2) | 0.7 (0.3, 1.4) |
| 5 | 100 | 1 | 1 | 1.1 (0.7, 1.6) | 1.1 (0.4, 8.8) | 1.8 (0.6, 9.1) | 0.4 (0.1, 1) | 1.1 (0.4, 8.7) | 1.5 (0.5, 9.1) | 0.9 (0.3, 2.2) | 0.6 (0.3, 0.9) | 0.4 (0.2, 0.7) | 0.5 (0.1, 0.8) |
| 5 | 250 | 1 | 0 | 1.1 (0.8, 1.4) | 0.9 (0.4, 2.1) | 1.3 (0.8, 3) | 1.2 (0.7, 4.2) | 0.9 (0.4, 2.1) | 1.2 (0.7, 2.7) | 1.1 (0.7, 2.3) | 1 (0.7, 1.5) | 0.8 (0.6, 1.3) | 0.9 (0.6, 1.4) |
| 5 | 250 | 1 | 1 | 1.1 (0.9, 1.4) | 1.1 (0.7, 2.4) | 1.4 (0.8, 3.9) | 0.7 (0.4, 1.2) | 1.1 (0.7, 2.4) | 1.2 (0.7, 3.5) | 1.1 (0.7, 2) | 0.7 (0.5, 1) | 0.6 (0.4, 0.9) | 0.7 (0.5, 1) |
| 5 | 500 | 1 | 0 | 1 (0.8, 1.2) | 0.9 (0.5, 1.4) | 1 (0.6, 1.6) | 1.1 (0.8, 2.8) | 0.9 (0.5, 1.4) | 0.9 (0.5, 1.5) | 0.9 (0.7, 1.5) | 0.9 (0.7, 1.2) | 0.9 (0.7, 1.2) | 0.9 (0.7, 1.2) |
| 5 | 500 | 1 | 1 | 1 (0.9, 1.2) | 0.9 (0.5, 1.4) | 1.2 (0.8, 1.8) | 0.8 (0.6, 1.2) | 0.9 (0.5, 1.4) | 1 (0.7, 1.6) | 1 (0.7, 1.5) | 0.8 (0.6, 1) | 0.7 (0.6, 1) | 0.8 (0.6, 1) |
| 10 | 100 | 1 | 0 | 1 (0.6, 1.5) | 1.1 (0.5, 4.3) | 1.4 (0.5, 8) | 0.4 (0, 1.2) | 1.1 (0.5, 4.7) | 1.3 (0.6, 9.1) | 0.9 (0.1, 2) | 0.8 (0.5, 1.1) | 0.5 (0.3, 0.9) | 0.6 (0.3, 1) |
| 10 | 100 | 1 | 1 | 1 (0.6, 1.6) | 1.1 (0.4, 4.8) | 1.5 (0.6, 7.1) | 0.2 (0, 1) | 1 (0.4, 4.9) | 1.3 (0.6, 7.6) | 0.8 (0, 1.8) | 0.6 (0.4, 0.9) | 0.4 (0.2, 0.6) | 0.5 (0.2, 0.8) |
| 10 | 250 | 1 | 0 | 1 (0.8, 1.2) | 1 (0.6, 1.5) | 1.2 (0.8, 1.8) | 0.7 (0.5, 1.2) | 1 (0.7, 1.5) | 1.1 (0.7, 1.7) | 1 (0.6, 1.5) | 0.9 (0.7, 1.1) | 0.7 (0.5, 1) | 0.8 (0.5, 1.1) |
| 10 | 250 | 1 | 1 | 1 (0.8, 1.2) | 1 (0.7, 1.5) | 1.2 (0.8, 2) | 0.6 (0.4, 1) | 1 (0.7, 1.5) | 1.1 (0.7, 1.8) | 1 (0.6, 1.5) | 0.8 (0.6, 1) | 0.6 (0.4, 0.8) | 0.7 (0.5, 0.9) |
| 10 | 500 | 1 | 0 | 1 (0.8, 1.1) | 1 (0.7, 1.3) | 1.1 (0.8, 1.4) | 0.8 (0.6, 1.1) | 1 (0.7, 1.3) | 1 (0.8, 1.3) | 1 (0.7, 1.3) | 0.9 (0.7, 1.1) | 0.8 (0.7, 1) | 0.9 (0.7, 1.1) |
| 10 | 500 | 1 | 1 | 1 (0.9, 1.1) | 1 (0.8, 1.3) | 1.1 (0.8, 1.5) | 0.8 (0.6, 1) | 1 (0.8, 1.3) | 1 (0.8, 1.4) | 1 (0.7, 1.3) | 0.8 (0.7, 1) | 0.8 (0.6, 1) | 0.8 (0.6, 1) |
| 2 | 100 | 0.5 | 0 | 1.2 (−0.4, 2.2) | 0.8 (0, 27.2) | 1.9 (−0.1, 27.2) | 3.6 (−0.2, 22.2) | 0.8 (0, 27.2) | 2.3 (0, 27.2) | 1.3 (−0.1, 18.1) | 0.9 (0, 2.5) | 0.6 (0, 2) | 0.8 (−0.2, 2.3) |
| 2 | 100 | 0.5 | 1 | 1.2 (−0.9, 1.8) | 1.2 (−0.4, 7.8) | 2 (−0.4, 7.8) | 0.3 (0, 0.9) | 1.2 (−0.4, 7.8) | 2.3 (−0.6, 7.8) | 0.7 (−0.2, 2.8) | 0.4 (0, 0.7) | 0.3 (0, 0.5) | 0.3 (0, 0.6) |
| 2 | 250 | 0.5 | 0 | 1.1 (0.7, 1.9) | 1.2 (0.1, 10.7) | 1.5 (0.3, 10.7) | 4.1 (0.8, 8) | 1.2 (0.1, 10.7) | 1.4 (0.1, 10.7) | 1.3 (0.3, 6.5) | 0.9 (0.5, 2) | 0.8 (0.4, 1.9) | 0.9 (0.5, 2.1) |
| 2 | 250 | 0.5 | 1 | 1.1 (0.8, 1.4) | 1.1 (0.4, 4.3) | 1.8 (0.6, 4.4) | 0.5 (0.2, 1.5) | 1.1 (0.4, 4.3) | 1.6 (0.5, 4.4) | 1 (0.4, 2.9) | 0.5 (0.2, 0.9) | 0.4 (0.2, 0.8) | 0.5 (0.2, 0.8) |
| 2 | 500 | 0.5 | 0 | 1.1 (0.8, 1.6) | 1.1 (0.6, 2.9) | 1.3 (0.7, 4) | 2.8 (1.1, 5) | 1.1 (0.6, 2.8) | 1.2 (0.7, 4.2) | 1.1 (0.6, 2.9) | 0.9 (0.6, 1.7) | 0.9 (0.6, 1.6) | 0.9 (0.6, 1.7) |
| 2 | 500 | 0.5 | 1 | 1.1 (0.8, 1.3) | 1.1 (0.5, 2.8) | 1.5 (0.7, 3.1) | 0.7 (0.4, 1.8) | 1 (0.5, 2.8) | 1.3 (0.6, 3.1) | 1.1 (0.6, 2.3) | 0.6 (0.4, 1) | 0.6 (0.4, 0.9) | 0.6 (0.4, 1) |
| 5 | 100 | 0.5 | 0 | 1.1 (0.3, 1.7) | 1.2 (0.1, 10.6) | 2 (0.1, 10.7) | 0.5 (0.1, 2.3) | 1.1 (0.1, 10.6) | 2.3 (0.2, 10.7) | 0.9 (0.1, 4) | 0.5 (0.1, 1) | 0.4 (0, 0.8) | 0.4 (0, 0.9) |
| 5 | 100 | 0.5 | 1 | 1.1 (0.1, 1.7) | 1 (0.1, 5.6) | 1.9 (0.1, 5.7) | 0.2 (0, 0.6) | 1 (0.1, 5.6) | 1.9 (0.1, 5.8) | 0.6 (0, 1.9) | 0.3 (0, 0.5) | 0.2 (0, 0.4) | 0.2 (0, 0.5) |
| 5 | 250 | 0.5 | 0 | 1.1 (0.8, 1.5) | 1.1 (0.1, 6.3) | 1.7 (0.7, 6.5) | 1 (0.4, 3.3) | 1.1 (0.1, 6.3) | 1.5 (0.6, 6.5) | 1.2 (0.5, 3.7) | 0.7 (0.4, 1.2) | 0.6 (0.3, 1.1) | 0.6 (0.4, 1.2) |
| 5 | 250 | 0.5 | 1 | 1.2 (0.8, 1.5) | 1.1 (0.4, 4) | 1.9 (0.6, 4.1) | 0.5 (0.2, 0.9) | 1.1 (0.4, 4) | 1.7 (0.5, 4.1) | 1 (0.4, 2.4) | 0.5 (0.3, 0.7) | 0.4 (0.2, 0.7) | 0.4 (0.2, 0.7) |
| 5 | 500 | 0.5 | 0 | 1.2 (0.9, 1.5) | 1.2 (0.7, 2.8) | 1.5 (0.8, 3.9) | 1.2 (0.7, 3.3) | 1.1 (0.6, 2.7) | 1.3 (0.7, 3.9) | 1.2 (0.7, 2.5) | 0.9 (0.6, 1.3) | 0.8 (0.5, 1.2) | 0.8 (0.6, 1.3) |
| 5 | 500 | 0.5 | 1 | 1.2 (1, 1.5) | 1.2 (0.6, 2.9) | 1.7 (0.9, 3) | 0.7 (0.4, 1.3) | 1.1 (0.6, 2.9) | 1.4 (0.7, 3) | 1.2 (0.7, 2.3) | 0.7 (0.5, 1) | 0.6 (0.4, 0.9) | 0.6 (0.4, 0.9) |
| 10 | 100 | 0.5 | 0 | 1 (0.3, 1.9) | 1.1 (0.2, 7.3) | 1.8 (0.3, 7.3) | 0.3 (0, 1.1) | 1.1 (0.2, 7.3) | 1.8 (0.3, 7.5) | 0.8 (0.1, 2.1) | 0.5 (0, 0.7) | 0.3 (0, 0.6) | 0.4 (0, 0.7) |
| 10 | 100 | 0.5 | 1 | 1 (0.3, 2) | 1.1 (0.1, 5.8) | 1.9 (0.3, 5.9) | 0.2 (0, 0.7) | 1.1 (0.2, 5.8) | 1.8 (0.3, 5.9) | 0.7 (0, 1.7) | 0.4 (0, 0.6) | 0.2 (0, 0.4) | 0.3 (0, 0.5) |
| 10 | 250 | 0.5 | 0 | 1 (0.8, 1.3) | 1.1 (0.6, 3.6) | 1.5 (0.8, 4.1) | 0.6 (0.3, 1.3) | 1.1 (0.6, 3.6) | 1.3 (0.7, 4.1) | 1.1 (0.6, 2.1) | 0.6 (0.4, 0.9) | 0.6 (0.3, 0.8) | 0.6 (0.4, 0.9) |
| 10 | 250 | 0.5 | 1 | 1 (0.8, 1.3) | 1.1 (0.6, 3.4) | 1.6 (0.8, 3.6) | 0.5 (0.2, 1) | 1.1 (0.6, 3.4) | 1.3 (0.7, 3.6) | 1 (0.6, 1.9) | 0.5 (0.3, 0.8) | 0.5 (0.3, 0.7) | 0.5 (0.3, 0.7) |
| 10 | 500 | 0.5 | 0 | 1 (0.9, 1.2) | 1 (0.7, 1.6) | 1.3 (0.8, 2.2) | 0.8 (0.6, 1.3) | 1 (0.7, 1.6) | 1.1 (0.8, 1.9) | 1.1 (0.7, 1.6) | 0.8 (0.6, 1) | 0.7 (0.5, 1) | 0.8 (0.6, 1) |
| 10 | 500 | 0.5 | 1 | 1 (0.9, 1.2) | 1 (0.7, 1.7) | 1.3 (0.9, 2.3) | 0.7 (0.5, 1.1) | 1 (0.7, 1.7) | 1.1 (0.8, 2) | 1.1 (0.7, 1.6) | 0.7 (0.5, 0.9) | 0.6 (0.5, 0.9) | 0.7 (0.5, 0.9) |

Table S8. Simulation results showing median calibration slopes (with 5th and 95th percentile) across simulation scenarios with marginal event rate $E(Y) = 0.25$ that differed by the number of predictors $K \in \{2, 5, 10\}$, the sample size $N \in \{100, 250, 500\}$, the effect multiplier $a \in \{0.5, 1\}$ and noise absent (0) or present (1). OP, prediction oracle; D, deviance; GCV, generalized cross-validation; CE, classification error; RCV50, repeated 10-fold cross-validated deviance with $\theta = 0.5$; RCV95, repeated 10-fold cross-validated deviance with $\theta = 0.95$; AIC, Akaike's information criterion; IP, shrinkage based on informative priors; WP, shrinkage based on weakly informative priors; FLIC, Firth's logistic regression with intercept-correction.

| K | N | a | Noise | OP | D | GCV | CE | RCV50 | RCV95 | AIC | IP | WP | FLIC |
|---|---|---|---|---|---|---|---|---|---|---|---|---|---|
| 2 | 100 | 1 | 0 | 1.1 (0.2, 1.6) | 0.9 (0.2, 2.3) | 1.1 (0.6, 3.1) | 5.5 (1, 9.7) | 0.9 (0.2, 2.2) | 0.9 (0.2, 2.6) | 0.9 (0.6, 2.3) | 1.1 (0.8, 1.8) | 0.9 (0.7, 1.6) | 1 (0.7, 1.7) |
| 2 | 100 | 1 | 1 | 1.1 (0.9, 1.4) | 1.1 (0.6, 2.7) | 1.8 (0.9, 5.4) | 1.1 (0.5, 3.1) | 1.1 (0.6, 2.6) | 1.3 (0.7, 4.2) | 1.1 (0.6, 2.2) | 0.8 (0.5, 1.1) | 0.6 (0.4, 1) | 0.7 (0.4, 1.1) |
| 2 | 250 | 1 | 0 | 1.1 (0.9, 1.4) | 1.1 (0.3, 1.6) | 1.2 (0.5, 1.8) | 3.1 (1.2, 4.5) | 1.1 (0.3, 1.6) | 1.1 (0.3, 1.7) | 1.1 (0.6, 1.6) | 1.1 (0.8, 1.5) | 1 (0.7, 1.4) | 1 (0.8, 1.4) |
| 2 | 250 | 1 | 1 | 1.1 (0.9, 1.3) | 1 (0.3, 1.6) | 1.3 (0.9, 2.4) | 1.3 (0.8, 2.8) | 1 (0.3, 1.6) | 1.1 (0.7, 1.8) | 1.1 (0.6, 1.7) | 0.9 (0.7, 1.2) | 0.8 (0.6, 1.2) | 0.9 (0.7, 1.2) |
| 2 | 500 | 1 | 0 | 1 (0.9, 1.2) | 1 (0.8, 1.3) | 1.1 (0.8, 1.4) | 2.1 (1.4, 2.6) | 1 (0.8, 1.3) | 1.1 (0.8, 1.4) | 1 (0.8, 1.3) | 1 (0.8, 1.3) | 1 (0.8, 1.2) | 1 (0.8, 1.2) |
| 2 | 500 | 1 | 1 | 1 (0.9, 1.2) | 1 (0.8, 1.3) | 1.2 (0.8, 1.6) | 1.2 (0.8, 2.1) | 1 (0.8, 1.3) | 1 (0.8, 1.4) | 1 (0.8, 1.4) | 0.9 (0.7, 1.1) | 0.9 (0.7, 1.1) | 0.9 (0.7, 1.1) |
| 5 | 100 | 1 | 0 | 1 (0.8, 1.3) | 1 (0.5, 2.1) | 1.5 (0.8, 3.6) | 1.3 (0.6, 5.2) | 1 (0.6, 2.1) | 1.2 (0.7, 2.6) | 1.1 (0.6, 2.1) | 0.9 (0.6, 1.3) | 0.7 (0.5, 1.1) | 0.8 (0.6, 1.2) |
| 5 | 100 | 1 | 1 | 1.1 (0.9, 1.3) | 1.1 (0.6, 2.4) | 1.8 (0.9, 5) | 0.8 (0.4, 2) | 1 (0.6, 2.3) | 1.3 (0.7, 3.5) | 1 (0.5, 1.9) | 0.7 (0.5, 1) | 0.5 (0.4, 0.8) | 0.6 (0.4, 0.9) |
| 5 | 250 | 1 | 0 | 1 (0.9, 1.2) | 1 (0.3, 1.4) | 1.2 (0.7, 1.8) | 1.4 (0.8, 3.5) | 1 (0.3, 1.4) | 1 (0.3, 1.5) | 1 (0.6, 1.5) | 0.9 (0.7, 1.2) | 0.9 (0.7, 1.2) | 0.9 (0.7, 1.2) |
| 5 | 250 | 1 | 1 | 1 (0.9, 1.2) | 1 (0.6, 1.5) | 1.4 (0.9, 2.2) | 1 (0.6, 2) | 1 (0.7, 1.5) | 1.1 (0.8, 1.6) | 1 (0.7, 1.5) | 0.8 (0.6, 1.1) | 0.7 (0.6, 1) | 0.8 (0.6, 1.1) |
| 5 | 500 | 1 | 0 | 1 (0.9, 1.2) | 1 (0.7, 1.3) | 1.1 (0.8, 1.5) | 1.4 (0.9, 2.4) | 1 (0.7, 1.3) | 1 (0.7, 1.3) | 1 (0.7, 1.3) | 1 (0.8, 1.2) | 0.9 (0.7, 1.2) | 0.9 (0.8, 1.2) |
| 5 | 500 | 1 | 1 | 1 (0.9, 1.1) | 1 (0.7, 1.3) | 1.2 (0.9, 1.6) | 1 (0.8, 1.7) | 1 (0.7, 1.3) | 1 (0.7, 1.3) | 1 (0.7, 1.3) | 0.9 (0.7, 1.1) | 0.9 (0.7, 1.1) | 0.9 (0.7, 1.1) |
| 10 | 100 | 1 | 0 | 1 (0.8, 1.2) | 1 (0.6, 1.8) | 1.5 (0.8, 3.4) | 0.9 (0.4, 2.3) | 1 (0.6, 1.8) | 1.2 (0.7, 2.2) | 1 (0.5, 1.7) | 0.8 (0.6, 1.1) | 0.6 (0.4, 0.9) | 0.7 (0.4, 1) |
| 10 | 100 | 1 | 1 | 1 (0.8, 1.3) | 1 (0.6, 1.9) | 1.7 (0.9, 4) | 0.7 (0.3, 2) | 1 (0.6, 1.9) | 1.2 (0.7, 2.4) | 0.9 (0.4, 1.6) | 0.7 (0.5, 0.9) | 0.5 (0.3, 0.7) | 0.6 (0.3, 0.9) |
| 10 | 250 | 1 | 0 | 1 (0.9, 1.2) | 1 (0.7, 1.4) | 1.2 (0.9, 1.8) | 1 (0.7, 1.7) | 1 (0.7, 1.4) | 1.1 (0.8, 1.5) | 1 (0.7, 1.4) | 0.9 (0.7, 1.1) | 0.8 (0.6, 1.1) | 0.9 (0.7, 1.1) |
| 10 | 250 | 1 | 1 | 1 (0.9, 1.2) | 1 (0.7, 1.4) | 1.3 (0.9, 2) | 0.9 (0.6, 1.6) | 1 (0.7, 1.4) | 1.1 (0.8, 1.5) | 1 (0.7, 1.4) | 0.8 (0.7, 1) | 0.7 (0.6, 0.9) | 0.8 (0.6, 1) |
| 10 | 500 | 1 | 0 | 1 (0.9, 1.1) | 1 (0.9, 1.2) | 1.1 (0.9, 1.4) | 1 (0.8, 1.6) | 1 (0.8, 1.2) | 1.1 (0.9, 1.3) | 1 (0.8, 1.3) | 1 (0.8, 1.1) | 0.9 (0.8, 1.1) | 0.9 (0.8, 1.1) |
| 10 | 500 | 1 | 1 | 1 (0.9, 1.1) | 1 (0.8, 1.2) | 1.2 (0.9, 1.5) | 1 (0.8, 1.4) | 1 (0.8, 1.2) | 1.1 (0.9, 1.3) | 1 (0.8, 1.3) | 0.9 (0.8, 1.1) | 0.9 (0.7, 1) | 0.9 (0.8, 1.1) |
| 2 | 100 | 0.5 | 0 | 1.2 (0.9, 1.9) | 1.4 (0.1, 13.6) | 2.1 (0.3, 13.6) | 4.4 (0.9, 9.5) | 1.3 (0.1, 13.6) | 1.7 (0.1, 13.6) | 1.4 (0.3, 8.7) | 1 (0.6, 2.2) | 0.9 (0.4, 2) | 0.9 (0.5, 2.2) |
| 2 | 100 | 0.5 | 1 | 1.2 (0.8, 1.5) | 1.3 (0.4, 5.4) | 2.5 (0.6, 5.4) | 0.6 (0.2, 2) | 1.3 (0.4, 5.4) | 2 (0.5, 5.4) | 1.1 (0.4, 3) | 0.5 (0.2, 0.9) | 0.4 (0.2, 0.8) | 0.5 (0.2, 0.8) |
| 2 | 250 | 0.5 | 0 | 1.1 (0.9, 1.7) | 1.2 (0.7, 2.9) | 1.5 (0.8, 4.6) | 2.9 (1.2, 5) | 1.2 (0.7, 2.8) | 1.4 (0.7, 4) | 1.3 (0.7, 2.9) | 1 (0.7, 1.7) | 1 (0.6, 1.7) | 1 (0.6, 1.7) |
| 2 | 250 | 0.5 | 1 | 1.1 (1, 1.4) | 1.2 (0.6, 3.1) | 1.9 (0.9, 3.4) | 1 (0.5, 2.3) | 1.2 (0.6, 3.2) | 1.4 (0.7, 3.4) | 1.2 (0.7, 2.7) | 0.7 (0.5, 1.1) | 0.7 (0.4, 1) | 0.7 (0.5, 1.1) |
| 2 | 500 | 0.5 | 0 | 1.1 (1, 1.5) | 1.1 (0.8, 1.8) | 1.3 (0.9, 2.3) | 2 (1.4, 2.9) | 1.1 (0.8, 1.8) | 1.2 (0.8, 2) | 1.2 (0.8, 1.9) | 1 (0.7, 1.5) | 1 (0.7, 1.5) | 1 (0.7, 1.5) |
| 2 | 500 | 0.5 | 1 | 1.1 (1, 1.3) | 1.1 (0.8, 1.9) | 1.5 (1, 2.4) | 1.2 (0.7, 2) | 1.1 (0.8, 1.9) | 1.2 (0.8, 2.2) | 1.2 (0.8, 1.9) | 0.8 (0.6, 1.1) | 0.8 (0.6, 1.1) | 0.8 (0.6, 1.2) |
| 5 | 100 | 0.5 | 0 | 1 (0.8, 1.4) | 1.1 (0.3, 6.3) | 2 (0.6, 6.4) | 0.9 (0.4, 3.3) | 1 (0.4, 6.3) | 1.5 (0.5, 6.4) | 1 (0.4, 3.2) | 0.6 (0.3, 1) | 0.5 (0.3, 0.9) | 0.6 (0.3, 1) |
| 5 | 100 | 0.5 | 1 | 1 (0.8, 1.4) | 1.1 (0.4, 3.9) | 2.1 (0.7, 4) | 0.5 (0.2, 1.4) | 1 (0.4, 3.9) | 1.5 (0.5, 4) | 0.8 (0.4, 2) | 0.4 (0.2, 0.6) | 0.3 (0.1, 0.6) | 0.4 (0.1, 0.6) |
| 5 | 250 | 0.5 | 0 | 1 (0.9, 1.2) | 1 (0.6, 2.4) | 1.5 (0.8, 3.5) | 1.2 (0.6, 2.9) | 1 (0.6, 2.4) | 1.2 (0.6, 3.4) | 1.1 (0.6, 2.3) | 0.8 (0.5, 1.2) | 0.7 (0.5, 1.1) | 0.8 (0.5, 1.2) |
| 5 | 250 | 0.5 | 1 | 1 (0.9, 1.2) | 1 (0.6, 2.4) | 1.7 (0.9, 2.6) | 0.7 (0.4, 1.6) | 1 (0.6, 2.4) | 1.2 (0.6, 2.6) | 1 (0.6, 1.9) | 0.6 (0.4, 0.8) | 0.6 (0.4, 0.8) | 0.6 (0.4, 0.8) |
| 5 | 500 | 0.5 | 0 | 1.1 (0.9, 1.2) | 1.1 (0.8, 1.6) | 1.3 (0.9, 2.3) | 1.4 (0.8, 2.3) | 1 (0.8, 1.6) | 1.1 (0.8, 1.8) | 1.1 (0.8, 1.7) | 0.9 (0.7, 1.2) | 0.9 (0.7, 1.2) | 0.9 (0.7, 1.2) |
| 5 | 500 | 0.5 | 1 | 1.1 (1, 1.2) | 1 (0.7, 1.7) | 1.5 (1, 2.1) | 0.9 (0.6, 1.6) | 1 (0.8, 1.7) | 1.1 (0.8, 1.9) | 1.1 (0.8, 1.7) | 0.8 (0.6, 1) | 0.8 (0.6, 1) | 0.8 (0.6, 1) |
| 10 | 100 | 0.5 | 0 | 1 (0.7, 1.3) | 1 (0.5, 4.4) | 2 (0.8, 4.7) | 0.6 (0.3, 1.9) | 1 (0.5, 4.3) | 1.4 (0.6, 4.7) | 0.9 (0.4, 2.1) | 0.5 (0.3, 0.8) | 0.4 (0.2, 0.7) | 0.5 (0.2, 0.8) |
| 10 | 100 | 0.5 | 1 | 1 (0.7, 1.3) | 1 (0.5, 3.6) | 2.1 (0.8, 3.7) | 0.4 (0.2, 1.4) | 1 (0.4, 3.5) | 1.4 (0.5, 3.7) | 0.8 (0.4, 1.7) | 0.4 (0.2, 0.6) | 0.3 (0.1, 0.5) | 0.3 (0.2, 0.6) |
| 10 | 250 | 0.5 | 0 | 1.1 (0.9, 1.2) | 1.1 (0.7, 2) | 1.7 (1, 3.1) | 0.9 (0.6, 2) | 1.1 (0.7, 2) | 1.3 (0.8, 2.5) | 1.2 (0.7, 2) | 0.8 (0.6, 1.1) | 0.7 (0.5, 1) | 0.8 (0.6, 1.1) |
| 10 | 250 | 0.5 | 1 | 1.1 (0.9, 1.2) | 1.1 (0.7, 2.1) | 1.8 (1, 2.7) | 0.8 (0.5, 1.8) | 1.1 (0.7, 2.1) | 1.3 (0.8, 2.5) | 1.1 (0.7, 1.8) | 0.7 (0.5, 0.9) | 0.6 (0.4, 0.8) | 0.7 (0.5, 0.9) |
| 10 | 500 | 0.5 | 0 | 1 (0.9, 1.2) | 1 (0.8, 1.5) | 1.3 (1, 2) | 1 (0.7, 1.6) | 1 (0.8, 1.5) | 1.1 (0.8, 1.6) | 1.1 (0.8, 1.5) | 0.9 (0.7, 1.1) | 0.8 (0.7, 1.1) | 0.9 (0.7, 1.1) |
| 10 | 500 | 0.5 | 1 | 1 (0.9, 1.2) | 1 (0.8, 1.5) | 1.4 (1, 2) | 0.9 (0.7, 1.5) | 1 (0.8, 1.5) | 1.1 (0.8, 1.6) | 1.1 (0.8, 1.5) | 0.8 (0.6, 1) | 0.8 (0.6, 0.9) | 0.8 (0.6, 1) |

Table S9. Simulation results showing root mean squared distance of the logarithm of calibration slopes across simulation scenarios with marginal event rate $E(Y) = 0.10$ that differed by the number of predictors $K \in \{2, 5, 10\}$, the sample size $N \in \{100, 250, 500\}$, the effect multiplier $a \in \{0.5, 1\}$ and noise absent (0) or present (1). OP, prediction oracle; *D*, deviance; *GCV*, generalized cross-validation; *CE*, classification error; RCV50, repeated 10-fold cross-validated deviance with $\theta = 0.5$; RCV95, repeated 10-fold cross-validated deviance with $\theta = 0.95$; AIC, Akaike's information criterion; IP, shrinkage based on informative priors; WP, shrinkage based on weakly informative priors; FLIC, Firth's logistic regression with intercept-correction.

| K | N | a | Noise | OP | D | GCV | CE | RCV50 | RCV95 | AIC | IP | WP | FLIC |
|---|---|---|---|---|---|---|---|---|---|---|---|---|---|
| 2 | 100 | 1 | 0 | 0.57 | 0.92 | 1.08 | 1.54 | 0.89 | 1.24 | 0.83 | 0.61 | 0.56 | 0.59 |
| 2 | 100 | 1 | 1 | 0.77 | 1.1 | 1.25 | 0.9 | 1.1 | 1.32 | 0.73 | 0.69 | 0.88 | 0.85 |
| 2 | 250 | 1 | 0 | 0.28 | 0.36 | 0.42 | 1.32 | 0.35 | 0.42 | 0.37 | 0.3 | 0.27 | 0.29 |
| 2 | 250 | 1 | 1 | 0.37 | 0.47 | 0.61 | 0.43 | 0.46 | 0.57 | 0.4 | 0.34 | 0.41 | 0.39 |
| 2 | 500 | 1 | 0 | 0.19 | 0.22 | 0.24 | 0.95 | 0.21 | 0.23 | 0.22 | 0.2 | 0.19 | 0.19 |
| 2 | 500 | 1 | 1 | 0.24 | 0.26 | 0.31 | 0.33 | 0.25 | 0.28 | 0.26 | 0.23 | 0.26 | 0.25 |
| 5 | 100 | 1 | 0 | 0.48 | 1.08 | 1.22 | 0.7 | 1.08 | 1.29 | 0.72 | 0.52 | 0.6 | 0.6 |
| 5 | 100 | 1 | 1 | 0.73 | 1.03 | 1.18 | 1.08 | 1.03 | 1.23 | 0.72 | 0.74 | 0.97 | 0.92 |
| 5 | 250 | 1 | 0 | 0.23 | 0.37 | 0.49 | 0.54 | 0.36 | 0.45 | 0.37 | 0.25 | 0.28 | 0.27 |
| 5 | 250 | 1 | 1 | 0.38 | 0.41 | 0.58 | 0.45 | 0.41 | 0.52 | 0.35 | 0.38 | 0.47 | 0.43 |
| 5 | 500 | 1 | 0 | 0.18 | 0.21 | 0.25 | 0.39 | 0.21 | 0.23 | 0.22 | 0.18 | 0.19 | 0.19 |
| 5 | 500 | 1 | 1 | 0.25 | 0.25 | 0.31 | 0.27 | 0.24 | 0.25 | 0.24 | 0.25 | 0.28 | 0.26 |
| 10 | 100 | 1 | 0 | 0.41 | 0.71 | 0.87 | 0.6 | 0.7 | 0.94 | 0.49 | 0.3 | 0.46 | 0.39 |
| 10 | 100 | 1 | 1 | 0.46 | 0.7 | 0.89 | 0.91 | 0.7 | 0.92 | 0.74 | 0.42 | 0.63 | 0.51 |
| 10 | 250 | 1 | 0 | 0.15 | 0.2 | 0.28 | 0.23 | 0.2 | 0.24 | 0.21 | 0.15 | 0.21 | 0.19 |
| 10 | 250 | 1 | 1 | 0.2 | 0.22 | 0.33 | 0.31 | 0.21 | 0.27 | 0.22 | 0.2 | 0.29 | 0.24 |
| 10 | 500 | 1 | 0 | 0.11 | 0.11 | 0.14 | 0.13 | 0.11 | 0.13 | 0.12 | 0.1 | 0.13 | 0.12 |
| 10 | 500 | 1 | 1 | 0.13 | 0.12 | 0.17 | 0.17 | 0.12 | 0.14 | 0.13 | 0.13 | 0.17 | 0.14 |
| 2 | 100 | 0.5 | 0 | 1.21 | 1.91 | 1.97 | 1.79 | 1.89 | 2.09 | 1.63 | 1.27 | 1.31 | 1.4 |
| 2 | 100 | 0.5 | 1 | 1.54 | 1.7 | 1.69 | 1.91 | 1.7 | 1.72 | 1.49 | 1.73 | 1.92 | 1.94 |
| 2 | 250 | 0.5 | 0 | 0.55 | 1.01 | 1.12 | 1.43 | 1 | 1.16 | 0.91 | 0.59 | 0.59 | 0.61 |
| 2 | 250 | 0.5 | 1 | 0.75 | 0.86 | 0.91 | 0.94 | 0.85 | 0.91 | 0.69 | 0.91 | 1 | 1.01 |
| 2 | 500 | 0.5 | 0 | 0.32 | 0.54 | 0.63 | 1.05 | 0.53 | 0.63 | 0.54 | 0.33 | 0.33 | 0.34 |
| 2 | 500 | 0.5 | 1 | 0.43 | 0.52 | 0.62 | 0.52 | 0.52 | 0.59 | 0.44 | 0.52 | 0.56 | 0.55 |
| 5 | 100 | 0.5 | 0 | 1.08 | 1.61 | 1.65 | 1.36 | 1.61 | 1.73 | 1.29 | 1.35 | 1.5 | 1.53 |
| 5 | 100 | 0.5 | 1 | 1.27 | 1.48 | 1.44 | 2.02 | 1.48 | 1.49 | 1.36 | 1.76 | 1.99 | 1.96 |
| 5 | 250 | 0.5 | 0 | 0.36 | 0.85 | 0.95 | 0.62 | 0.84 | 0.95 | 0.65 | 0.57 | 0.63 | 0.63 |
| 5 | 250 | 0.5 | 1 | 0.61 | 0.76 | 0.8 | 0.99 | 0.75 | 0.81 | 0.6 | 0.94 | 1.05 | 1.02 |
| 5 | 500 | 0.5 | 0 | 0.19 | 0.43 | 0.58 | 0.5 | 0.42 | 0.54 | 0.41 | 0.32 | 0.35 | 0.34 |
| 5 | 500 | 0.5 | 1 | 0.37 | 0.48 | 0.56 | 0.56 | 0.47 | 0.52 | 0.38 | 0.56 | 0.61 | 0.59 |
| 10 | 100 | 0.5 | 0 | 0.89 | 1.24 | 1.33 | 1.3 | 1.23 | 1.38 | 0.92 | 1.08 | 1.26 | 1.07 |
| 10 | 100 | 0.5 | 1 | 0.91 | 1.16 | 1.23 | 1.67 | 1.15 | 1.26 | 1.14 | 1.27 | 1.51 | 1.29 |
| 10 | 250 | 0.5 | 0 | 0.31 | 0.57 | 0.73 | 0.48 | 0.57 | 0.7 | 0.4 | 0.43 | 0.5 | 0.44 |
| 10 | 250 | 0.5 | 1 | 0.35 | 0.57 | 0.72 | 0.66 | 0.57 | 0.68 | 0.37 | 0.58 | 0.68 | 0.6 |
| 10 | 500 | 0.5 | 0 | 0.18 | 0.28 | 0.41 | 0.29 | 0.28 | 0.34 | 0.28 | 0.25 | 0.28 | 0.26 |
| 10 | 500 | 0.5 | 1 | 0.22 | 0.3 | 0.45 | 0.35 | 0.29 | 0.36 | 0.27 | 0.34 | 0.38 | 0.34 |

Table S10. Simulation results showing root mean squared distance of the logarithm of calibration slopes across simulation scenarios with marginal event rate $E(Y) = 0.25$ that differed by the number of predictors $K \in \{2, 5, 10\}$, the sample size $N \in \{100, 250, 500\}$, the effect multiplier $a \in \{0.5, 1\}$ and noise absent (0) or present (1). OP, prediction oracle; *D*, deviance; *GCV*, generalized cross-validation; *CE*, classification error; RCV50, repeated 10-fold cross-validated deviance with $\theta = 0.5$; RCV95, repeated 10-fold cross-validated deviance with $\theta = 0.95$; AIC, Akaike's information criterion; IP, shrinkage based on informative priors; WP, shrinkage based on weakly informative priors; FLIC, Firth's logistic regression with intercept-correction.

| *K* | *N* | *a* | Noise | OP | *D* | *GCV* | *CE* | RCV50 | RCV95 | AIC | IP | WP | FLIC |
|---|---|---|---|---|---|---|---|---|---|---|---|---|---|
| 2 | 100 | 1 | 0 | 0.23 | 0.37 | 0.49 | 1.49 | 0.36 | 0.44 | 0.35 | 0.26 | 0.24 | 0.26 |
| 2 | 100 | 1 | 1 | 0.3 | 0.45 | 0.75 | 0.47 | 0.44 | 0.59 | 0.36 | 0.28 | 0.37 | 0.33 |
| 2 | 250 | 1 | 0 | 0.15 | 0.19 | 0.23 | 0.96 | 0.18 | 0.21 | 0.19 | 0.16 | 0.16 | 0.16 |
| 2 | 250 | 1 | 1 | 0.16 | 0.21 | 0.36 | 0.41 | 0.2 | 0.24 | 0.22 | 0.16 | 0.19 | 0.17 |
| 2 | 500 | 1 | 0 | 0.11 | 0.13 | 0.14 | 0.65 | 0.12 | 0.14 | 0.12 | 0.11 | 0.11 | 0.11 |
| 2 | 500 | 1 | 1 | 0.11 | 0.13 | 0.2 | 0.32 | 0.13 | 0.14 | 0.14 | 0.12 | 0.13 | 0.12 |
| 5 | 100 | 1 | 0 | 0.17 | 0.33 | 0.57 | 0.66 | 0.32 | 0.43 | 0.32 | 0.2 | 0.25 | 0.24 |
| 5 | 100 | 1 | 1 | 0.3 | 0.41 | 0.73 | 0.39 | 0.4 | 0.53 | 0.32 | 0.32 | 0.43 | 0.37 |
| 5 | 250 | 1 | 0 | 0.12 | 0.17 | 0.25 | 0.53 | 0.16 | 0.19 | 0.18 | 0.13 | 0.14 | 0.14 |
| 5 | 250 | 1 | 1 | 0.17 | 0.19 | 0.35 | 0.27 | 0.19 | 0.21 | 0.2 | 0.18 | 0.22 | 0.19 |
| 5 | 500 | 1 | 0 | 0.09 | 0.11 | 0.15 | 0.4 | 0.11 | 0.12 | 0.12 | 0.1 | 0.1 | 0.1 |
| 5 | 500 | 1 | 1 | 0.1 | 0.12 | 0.2 | 0.19 | 0.12 | 0.13 | 0.13 | 0.12 | 0.13 | 0.12 |
| 10 | 100 | 1 | 0 | 0.16 | 0.24 | 0.49 | 0.32 | 0.24 | 0.32 | 0.24 | 0.17 | 0.27 | 0.23 |
| 10 | 100 | 1 | 1 | 0.23 | 0.28 | 0.59 | 0.39 | 0.27 | 0.37 | 0.26 | 0.25 | 0.38 | 0.32 |
| 10 | 250 | 1 | 0 | 0.09 | 0.11 | 0.21 | 0.19 | 0.11 | 0.13 | 0.13 | 0.1 | 0.13 | 0.12 |
| 10 | 250 | 1 | 1 | 0.12 | 0.12 | 0.26 | 0.19 | 0.12 | 0.15 | 0.13 | 0.13 | 0.18 | 0.15 |
| 10 | 500 | 1 | 0 | 0.06 | 0.07 | 0.11 | 0.12 | 0.07 | 0.08 | 0.08 | 0.07 | 0.08 | 0.08 |
| 10 | 500 | 1 | 1 | 0.08 | 0.08 | 0.14 | 0.12 | 0.08 | 0.09 | 0.08 | 0.08 | 0.1 | 0.09 |
| 2 | 100 | 0.5 | 0 | 0.62 | 1.16 | 1.33 | 1.48 | 1.14 | 1.32 | 1.02 | 0.65 | 0.65 | 0.68 |
| 2 | 100 | 0.5 | 1 | 0.8 | 0.99 | 1.1 | 0.89 | 0.99 | 1.07 | 0.79 | 0.94 | 1.03 | 1.01 |
| 2 | 250 | 0.5 | 0 | 0.23 | 0.45 | 0.63 | 0.98 | 0.44 | 0.55 | 0.46 | 0.26 | 0.26 | 0.27 |
| 2 | 250 | 0.5 | 1 | 0.32 | 0.48 | 0.65 | 0.42 | 0.48 | 0.55 | 0.42 | 0.41 | 0.45 | 0.42 |
| 2 | 500 | 0.5 | 0 | 0.16 | 0.24 | 0.34 | 0.64 | 0.24 | 0.28 | 0.26 | 0.19 | 0.19 | 0.19 |
| 2 | 500 | 0.5 | 1 | 0.19 | 0.27 | 0.43 | 0.32 | 0.27 | 0.32 | 0.28 | 0.25 | 0.26 | 0.25 |
| 5 | 100 | 0.5 | 0 | 0.41 | 0.88 | 1.07 | 0.74 | 0.87 | 1.01 | 0.66 | 0.64 | 0.72 | 0.72 |
| 5 | 100 | 0.5 | 1 | 0.58 | 0.8 | 0.9 | 0.9 | 0.8 | 0.86 | 0.61 | 0.96 | 1.09 | 1.04 |
| 5 | 250 | 0.5 | 0 | 0.15 | 0.39 | 0.62 | 0.52 | 0.39 | 0.49 | 0.38 | 0.29 | 0.31 | 0.3 |
| 5 | 250 | 0.5 | 1 | 0.29 | 0.41 | 0.57 | 0.42 | 0.4 | 0.47 | 0.33 | 0.47 | 0.52 | 0.48 |
| 5 | 500 | 0.5 | 0 | 0.11 | 0.21 | 0.37 | 0.4 | 0.21 | 0.25 | 0.24 | 0.18 | 0.19 | 0.19 |
| 5 | 500 | 0.5 | 1 | 0.18 | 0.24 | 0.39 | 0.26 | 0.24 | 0.28 | 0.24 | 0.28 | 0.3 | 0.28 |
| 10 | 100 | 0.5 | 0 | 0.31 | 0.65 | 0.89 | 0.56 | 0.64 | 0.78 | 0.41 | 0.56 | 0.68 | 0.6 |
| 10 | 100 | 0.5 | 1 | 0.41 | 0.63 | 0.83 | 0.73 | 0.63 | 0.72 | 0.43 | 0.76 | 0.9 | 0.8 |
| 10 | 250 | 0.5 | 0 | 0.16 | 0.3 | 0.55 | 0.31 | 0.3 | 0.37 | 0.28 | 0.27 | 0.31 | 0.28 |
| 10 | 250 | 0.5 | 1 | 0.22 | 0.31 | 0.56 | 0.35 | 0.31 | 0.38 | 0.26 | 0.38 | 0.42 | 0.38 |
| 10 | 500 | 0.5 | 0 | 0.11 | 0.16 | 0.33 | 0.2 | 0.16 | 0.19 | 0.18 | 0.16 | 0.18 | 0.16 |
| 10 | 500 | 0.5 | 1 | 0.15 | 0.18 | 0.37 | 0.21 | 0.17 | 0.21 | 0.19 | 0.22 | 0.24 | 0.22 |

Figure S1. Scatter plots showing the logarithm of calibration slopes obtained by optimizing different tuning criteria versus tuned complexity parameter values $\lambda^*$ over 1000 generated datasets in scenarios with marginal event rate $\mathrm{E}(Y) = 0.10$, the number of predictors $K = 5$, noise absent or present, the sample size of $N \in \{100, 250, 500\}$ considering A) moderate ($a = 0.5$) and B) strong ($a = 1$) predictors. The horizontal line indicates the calibration slope of 1. Red points refer to datasets where separation occurred. OP, prediction oracle; D, deviance; GCV, generalized cross-validation; CE, classification error; RCV50, repeated 10-fold cross-validated deviance with $\theta = 0.5$; RCV95, repeated 10-fold cross-validated deviance with $\theta = 0.95$; AIC, Akaike's information criterion.

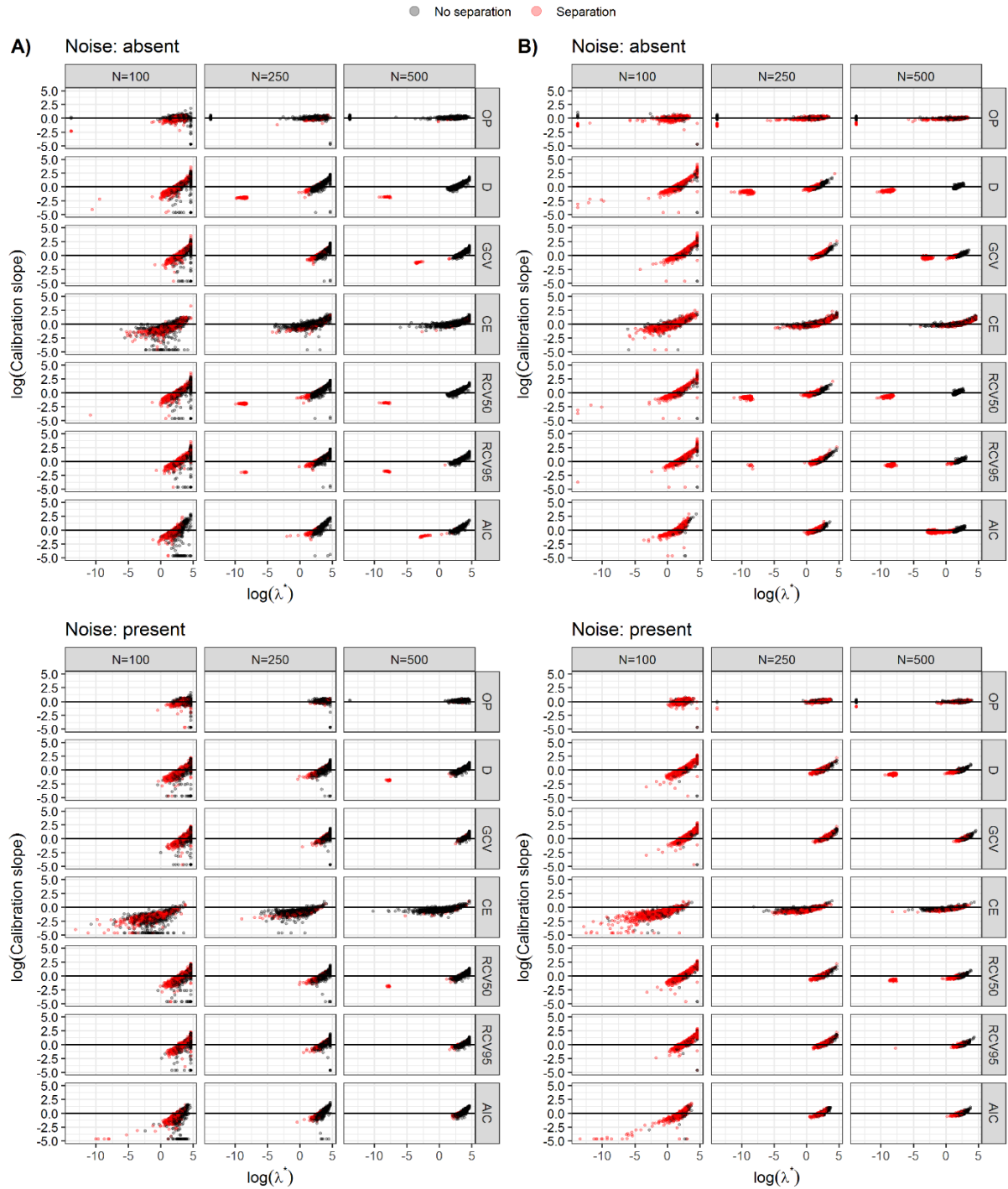

Table S11. Simulation results showing mean c-indices (with standard deviations) (× 1000) across simulation scenarios with marginal event rate with marginal event rate $E(Y) = 0.10$ that differed by the number of predictors $K \in \{2, 5, 10\}$, the sample size $N \in \{100, 250, 500\}$, the effect multiplier $a \in \{0.5, 1\}$ and noise absent (0) or present (1). Optimal c-index was calculated based on event probabilities from the true model. OP, prediction oracle; D, deviance; GCV, generalized cross-validation; CE, classification error; RCV50, repeated 10-fold cross-validated deviance with $\theta = 0.5$; RCV95, repeated 10-fold cross-validated deviance with $\theta = 0.95$; AIC, Akaike's information criterion; IP, shrinkage based on informative priors; WP, shrinkage based on weakly informative priors; FLIC, Firth's logistic regression with intercept-correction.

| K | N | a | Noise | Optimal | OP | D | GCV | CE | RCV50 | RCV95 | AIC | IP | WP | FLIC |
|---|---|---|---|---|---|---|---|---|---|---|---|---|---|---|
| 2 | 100 | 1 | 0 | 724 (70) | 716 (75) | 717 (74) | 717 (74) | 717 (74) | 717 (75) | 717 (74) | 716 (75) | 716 (75) | 715 (75) | 712 (76) |
| 2 | 100 | 1 | 1 | 724 (70) | 691 (86) | 691 (86) | 692 (86) | 687 (87) | 691 (86) | 692 (86) | 690 (86) | 689 (87) | 686 (87) | 678 (89) |
| 2 | 250 | 1 | 0 | 726 (44) | 724 (45) | 724 (45) | 724 (45) | 722 (44) | 724 (45) | 724 (45) | 724 (45) | 723 (45) | 724 (45) | 723 (45) |
| 2 | 250 | 1 | 1 | 726 (44) | 712 (52) | 713 (52) | 713 (52) | 712 (52) | 713 (52) | 713 (52) | 713 (52) | 712 (52) | 711 (52) | 709 (53) |
| 2 | 500 | 1 | 0 | 727 (30) | 726 (31) | 726 (31) | 726 (31) | 724 (31) | 726 (31) | 726 (31) | 726 (31) | 726 (31) | 726 (31) | 726 (31) |
| 2 | 500 | 1 | 1 | 727 (30) | 723 (35) | 722 (36) | 722 (36) | 722 (36) | 722 (36) | 722 (36) | 722 (36) | 723 (36) | 723 (35) | 722 (36) |
| 5 | 100 | 1 | 0 | 766 (74) | 723 (88) | 724 (88) | 724 (88) | 722 (89) | 724 (88) | 724 (88) | 723 (88) | 723 (88) | 722 (88) | 715 (90) |
| 5 | 100 | 1 | 1 | 766 (74) | 697 (89) | 698 (89) | 699 (89) | 692 (89) | 698 (89) | 699 (89) | 696 (89) | 694 (88) | 691 (88) | 680 (89) |
| 5 | 250 | 1 | 0 | 762 (45) | 744 (48) | 743 (48) | 744 (49) | 744 (49) | 744 (48) | 744 (49) | 744 (48) | 744 (48) | 743 (48) | 742 (49) |
| 5 | 250 | 1 | 1 | 762 (45) | 728 (52) | 728 (52) | 728 (52) | 726 (52) | 728 (52) | 728 (52) | 728 (52) | 727 (52) | 726 (52) | 723 (53) |
| 5 | 500 | 1 | 0 | 763 (33) | 754 (34) | 754 (34) | 754 (34) | 754 (34) | 754 (34) | 754 (34) | 754 (34) | 755 (34) | 755 (34) | 754 (34) |
| 5 | 500 | 1 | 1 | 763 (33) | 745 (35) | 744 (35) | 744 (35) | 744 (35) | 744 (35) | 744 (35) | 745 (35) | 745 (35) | 744 (35) | 744 (35) |
| 10 | 100 | 1 | 0 | 850 (65) | 794 (86) | 795 (86) | 796 (86) | 783 (92) | 795 (86) | 796 (86) | 792 (88) | 792 (88) | 787 (90) | 782 (90) |
| 10 | 100 | 1 | 1 | 850 (65) | 780 (92) | 781 (91) | 782 (91) | 761 (96) | 781 (91) | 782 (91) | 771 (95) | 776 (93) | 768 (95) | 758 (96) |
| 10 | 250 | 1 | 0 | 848 (41) | 821 (46) | 820 (46) | 820 (46) | 818 (47) | 820 (46) | 820 (46) | 820 (46) | 821 (46) | 820 (47) | 818 (47) |
| 10 | 250 | 1 | 1 | 848 (41) | 811 (48) | 810 (48) | 810 (48) | 806 (49) | 810 (48) | 811 (48) | 810 (48) | 809 (48) | 807 (48) | 805 (49) |
| 10 | 500 | 1 | 0 | 850 (29) | 836 (31) | 836 (31) | 835 (31) | 836 (31) | 836 (31) | 836 (31) | 836 (31) | 837 (31) | 836 (31) | 836 (31) |
| 10 | 500 | 1 | 1 | 850 (29) | 829 (33) | 829 (33) | 828 (33) | 828 (33) | 829 (33) | 829 (33) | 829 (33) | 829 (33) | 828 (33) | 828 (33) |
| 2 | 100 | 0.5 | 0 | 635 (75) | 621 (75) | 623 (74) | 624 (74) | 623 (75) | 623 (75) | 624 (74) | 624 (74) | 621 (75) | 619 (75) | 618 (75) |
| 2 | 100 | 0.5 | 1 | 635 (75) | 608 (74) | 609 (74) | 609 (75) | 604 (73) | 609 (74) | 609 (75) | 608 (74) | 605 (74) | 604 (74) | 602 (73) |
| 2 | 250 | 0.5 | 0 | 637 (50) | 629 (56) | 630 (55) | 631 (55) | 631 (54) | 630 (55) | 631 (55) | 631 (55) | 629 (57) | 629 (57) | 628 (56) |
| 2 | 250 | 0.5 | 1 | 637 (50) | 611 (58) | 611 (58) | 612 (58) | 607 (58) | 611 (58) | 612 (58) | 611 (58) | 607 (58) | 606 (58) | 604 (57) |
| 2 | 500 | 0.5 | 0 | 634 (38) | 631 (40) | 631 (40) | 631 (40) | 631 (39) | 631 (40) | 631 (40) | 631 (40) | 631 (40) | 631 (40) | 631 (40) |
| 2 | 500 | 0.5 | 1 | 634 (38) | 619 (44) | 619 (44) | 619 (44) | 617 (45) | 619 (44) | 619 (44) | 619 (44) | 616 (45) | 616 (45) | 615 (45) |
| 5 | 100 | 0.5 | 0 | 655 (80) | 615 (76) | 616 (76) | 616 (77) | 614 (76) | 616 (76) | 616 (77) | 616 (76) | 614 (76) | 613 (76) | 610 (75) |
| 5 | 100 | 0.5 | 1 | 655 (80) | 608 (76) | 608 (77) | 608 (77) | 603 (75) | 608 (77) | 608 (77) | 607 (76) | 604 (75) | 603 (74) | 598 (73) |
| 5 | 250 | 0.5 | 0 | 655 (53) | 627 (57) | 627 (57) | 628 (57) | 626 (57) | 627 (57) | 628 (57) | 628 (57) | 625 (58) | 624 (58) | 623 (58) |
| 5 | 250 | 0.5 | 1 | 655 (53) | 615 (57) | 615 (57) | 616 (57) | 610 (57) | 615 (57) | 616 (57) | 614 (57) | 610 (57) | 609 (57) | 607 (57) |
| 5 | 500 | 0.5 | 0 | 657 (37) | 639 (42) | 639 (42) | 640 (42) | 639 (42) | 639 (42) | 640 (42) | 639 (42) | 638 (43) | 637 (43) | 637 (43) |
| 5 | 500 | 0.5 | 1 | 657 (37) | 629 (43) | 629 (43) | 630 (43) | 626 (44) | 629 (43) | 630 (43) | 629 (43) | 626 (44) | 625 (44) | 624 (44) |
| 10 | 100 | 0.5 | 0 | 723 (88) | 660 (92) | 661 (92) | 662 (92) | 652 (90) | 661 (92) | 662 (93) | 659 (93) | 655 (91) | 651 (90) | 650 (90) |
| 10 | 100 | 0.5 | 1 | 723 (88) | 651 (88) | 651 (88) | 652 (89) | 637 (86) | 651 (88) | 652 (89) | 648 (87) | 643 (87) | 638 (86) | 636 (85) |
| 10 | 250 | 0.5 | 0 | 720 (56) | 679 (62) | 680 (62) | 681 (62) | 675 (63) | 680 (62) | 681 (62) | 680 (62) | 675 (63) | 674 (63) | 674 (62) |
| 10 | 250 | 0.5 | 1 | 720 (56) | 670 (64) | 670 (64) | 672 (64) | 662 (65) | 671 (64) | 671 (64) | 670 (64) | 663 (65) | 661 (65) | 660 (65) |
| 10 | 500 | 0.5 | 0 | 718 (37) | 694 (41) | 694 (41) | 695 (41) | 692 (42) | 694 (41) | 695 (41) | 694 (41) | 692 (42) | 692 (42) | 692 (42) |
| 10 | 500 | 0.5 | 1 | 718 (37) | 685 (42) | 686 (42) | 686 (42) | 681 (43) | 686 (42) | 686 (42) | 686 (42) | 681 (43) | 680 (43) | 680 (43) |

Table S12. Simulation results showing mean c-indices (with standard deviations) (× 1000) across simulation scenarios with marginal event rate with marginal event rate $E(Y) = 0.25$ that differed by the number of predictors $K \in \{2, 5, 10\}$, the sample size $N \in \{100, 250, 500\}$, the effect multiplier $a \in \{0.5, 1\}$ and noise absent (0) or present (1). Optimal c-index was calculated based on event probabilities from the true model. OP, prediction oracle; D, deviance; GCV, generalized cross-validation; CE, classification error; RCV50, repeated 10-fold cross-validated deviance with $\theta = 0.5$; RCV95, repeated 10-fold cross-validated deviance with $\theta = 0.95$; AIC, Akaike's information criterion; IP, shrinkage based on informative priors; WP, shrinkage based on weakly informative priors; FLIC, Firth's logistic regression with intercept-correction.

| K | N | a | Noise | Optimal | OP | D | GCV | CE | RCV50 | RCV95 | AIC | IP | WP | FLIC |
|---|---|---|---|---|---|---|---|---|---|---|---|---|---|---|
| 2 | 100 | 1 | 0 | 737 (50) | 735 (51) | 735 (51) | 734 (51) | 733 (51) | 735 (51) | 735 (51) | 735 (51) | 735 (51) | 735 (51) | 734 (51) |
| 2 | 100 | 1 | 1 | 737 (50) | 720 (62) | 721 (61) | 722 (61) | 719 (62) | 721 (61) | 722 (61) | 721 (61) | 718 (62) | 716 (63) | 715 (63) |
| 2 | 250 | 1 | 0 | 736 (31) | 735 (32) | 735 (31) | 735 (32) | 733 (31) | 735 (31) | 735 (31) | 735 (31) | 735 (31) | 735 (31) | 735 (31) |
| 2 | 250 | 1 | 1 | 736 (31) | 732 (36) | 731 (36) | 731 (36) | 731 (36) | 731 (36) | 731 (36) | 731 (36) | 731 (36) | 731 (36) | 731 (36) |
| 2 | 500 | 1 | 0 | 736 (22) | 735 (22) | 735 (22) | 735 (22) | 732 (23) | 735 (22) | 735 (23) | 735 (22) | 735 (22) | 735 (22) | 735 (22) |
| 2 | 500 | 1 | 1 | 736 (22) | 734 (25) | 734 (25) | 733 (25) | 733 (25) | 734 (25) | 734 (25) | 734 (25) | 734 (25) | 734 (25) | 734 (25) |
| 5 | 100 | 1 | 0 | 768 (50) | 746 (57) | 746 (57) | 746 (57) | 746 (57) | 746 (57) | 746 (57) | 746 (57) | 746 (57) | 745 (57) | 743 (58) |
| 5 | 100 | 1 | 1 | 768 (50) | 731 (59) | 731 (58) | 732 (58) | 728 (59) | 731 (59) | 731 (58) | 731 (59) | 727 (60) | 725 (60) | 722 (60) |
| 5 | 250 | 1 | 0 | 768 (33) | 760 (34) | 760 (34) | 760 (34) | 759 (35) | 760 (34) | 760 (34) | 760 (34) | 760 (34) | 760 (34) | 760 (34) |
| 5 | 250 | 1 | 1 | 768 (33) | 750 (35) | 750 (35) | 750 (35) | 749 (35) | 750 (35) | 750 (35) | 750 (35) | 749 (35) | 749 (35) | 748 (35) |
| 5 | 500 | 1 | 0 | 769 (24) | 765 (25) | 765 (24) | 765 (24) | 764 (24) | 765 (24) | 765 (24) | 765 (24) | 765 (24) | 765 (24) | 765 (24) |
| 5 | 500 | 1 | 1 | 769 (24) | 759 (25) | 759 (25) | 758 (25) | 758 (25) | 759 (25) | 759 (25) | 759 (25) | 759 (25) | 759 (25) | 759 (25) |
| 10 | 100 | 1 | 0 | 830 (47) | 794 (54) | 794 (54) | 793 (55) | 792 (55) | 794 (54) | 794 (55) | 794 (54) | 794 (54) | 792 (54) | 789 (55) |
| 10 | 100 | 1 | 1 | 830 (47) | 781 (56) | 780 (57) | 780 (57) | 777 (58) | 780 (57) | 780 (57) | 780 (57) | 778 (56) | 775 (57) | 771 (57) |
| 10 | 250 | 1 | 0 | 825 (28) | 808 (30) | 808 (31) | 807 (31) | 808 (31) | 808 (31) | 808 (31) | 808 (31) | 808 (31) | 808 (31) | 808 (31) |
| 10 | 250 | 1 | 1 | 825 (28) | 801 (32) | 801 (32) | 800 (32) | 800 (32) | 801 (32) | 800 (32) | 800 (32) | 800 (32) | 800 (32) | 799 (32) |
| 10 | 500 | 1 | 0 | 826 (19) | 818 (20) | 817 (20) | 817 (20) | 817 (20) | 817 (20) | 817 (20) | 817 (20) | 817 (20) | 817 (20) | 817 (20) |
| 10 | 500 | 1 | 1 | 826 (19) | 813 (20) | 813 (20) | 812 (21) | 812 (20) | 813 (20) | 813 (20) | 813 (20) | 813 (20) | 813 (20) | 812 (20) |
| 2 | 100 | 0.5 | 0 | 642 (57) | 631 (61) | 633 (60) | 633 (60) | 634 (60) | 633 (60) | 633 (60) | 633 (60) | 630 (61) | 629 (61) | 628 (61) |
| 2 | 100 | 0.5 | 1 | 642 (57) | 616 (63) | 616 (63) | 617 (63) | 613 (63) | 616 (63) | 617 (63) | 616 (63) | 611 (62) | 610 (62) | 608 (62) |
| 2 | 250 | 0.5 | 0 | 640 (37) | 638 (38) | 637 (38) | 638 (38) | 638 (37) | 637 (38) | 638 (38) | 637 (38) | 637 (39) | 637 (39) | 637 (39) |
| 2 | 250 | 0.5 | 1 | 640 (37) | 627 (42) | 627 (42) | 628 (42) | 625 (43) | 627 (42) | 627 (42) | 627 (42) | 623 (43) | 623 (43) | 623 (43) |
| 2 | 500 | 0.5 | 0 | 640 (25) | 639 (26) | 639 (26) | 639 (26) | 639 (25) | 639 (26) | 639 (26) | 639 (26) | 639 (26) | 639 (26) | 639 (26) |
| 2 | 500 | 0.5 | 1 | 640 (25) | 634 (29) | 634 (29) | 635 (28) | 634 (29) | 634 (29) | 635 (29) | 635 (29) | 633 (29) | 633 (29) | 633 (29) |
| 5 | 100 | 0.5 | 0 | 661 (60) | 630 (62) | 630 (63) | 632 (63) | 629 (63) | 631 (63) | 632 (63) | 631 (62) | 627 (62) | 626 (62) | 624 (63) |
| 5 | 100 | 0.5 | 1 | 661 (60) | 617 (62) | 618 (61) | 619 (62) | 612 (61) | 618 (61) | 618 (61) | 617 (61) | 611 (61) | 608 (61) | 607 (60) |
| 5 | 250 | 0.5 | 0 | 661 (39) | 645 (41) | 646 (41) | 647 (40) | 645 (41) | 646 (41) | 646 (41) | 646 (41) | 644 (41) | 643 (41) | 643 (41) |
| 5 | 250 | 0.5 | 1 | 661 (39) | 635 (42) | 635 (42) | 636 (42) | 632 (42) | 635 (42) | 635 (41) | 635 (42) | 631 (42) | 630 (42) | 629 (42) |
| 5 | 500 | 0.5 | 0 | 661 (27) | 654 (28) | 653 (28) | 654 (28) | 654 (28) | 653 (28) | 653 (28) | 653 (28) | 653 (28) | 653 (28) | 653 (28) |
| 5 | 500 | 0.5 | 1 | 661 (27) | 646 (28) | 646 (29) | 647 (28) | 645 (29) | 646 (29) | 646 (28) | 646 (28) | 644 (29) | 644 (29) | 644 (29) |
| 10 | 100 | 0.5 | 0 | 703 (62) | 653 (68) | 653 (68) | 654 (68) | 649 (68) | 653 (68) | 653 (68) | 652 (68) | 649 (69) | 647 (69) | 646 (69) |
| 10 | 100 | 0.5 | 1 | 703 (62) | 643 (68) | 644 (68) | 644 (68) | 638 (66) | 644 (68) | 644 (68) | 643 (68) | 637 (67) | 635 (66) | 633 (66) |
| 10 | 250 | 0.5 | 0 | 705 (39) | 680 (41) | 680 (41) | 680 (41) | 679 (42) | 680 (41) | 680 (41) | 681 (41) | 679 (42) | 679 (42) | 678 (42) |
| 10 | 250 | 0.5 | 1 | 705 (39) | 672 (42) | 672 (42) | 672 (42) | 669 (43) | 672 (42) | 672 (42) | 672 (42) | 669 (43) | 668 (43) | 667 (43) |
| 10 | 500 | 0.5 | 0 | 705 (27) | 690 (28) | 690 (28) | 689 (28) | 689 (28) | 690 (28) | 690 (28) | 690 (28) | 690 (28) | 689 (28) | 689 (28) |
| 10 | 500 | 0.5 | 1 | 705 (27) | 684 (28) | 683 (28) | 683 (29) | 683 (28) | 683 (28) | 683 (28) | 683 (28) | 682 (29) | 682 (29) | 682 (29) |